%% file: main.tex
\documentclass[a4paper,11pt]{article}
\pdfoutput=1 % if your are submitting a pdflatex (i.e. if you have
\usepackage{jcappub} % for details on the use of the package, please see the JCAP-author-manual
\usepackage[T1]{fontenc} % if needed
\usepackage{orcidlink}
\usepackage{booktabs}
\usepackage{multirow}

\title{\boldmath 
Examining the Linear and
Quadratic Dark Energy Parameterizations
with DESI DR2 via AIC and BIC Selection
}

 \author{Laiphangbam Khumaba Mangang$^{\dagger}$\orcidlink{0009-0009-2219-4583},}
 \author{Yumnam Prakash Singh$^\ast$\orcidlink{0009-0008-6329-8859},}
 \author{N. Chandrachani Devi$^\ddagger$\orcidlink{0000-0002-8939-7352}}
 \affiliation{Department of Physics, Manipur University, Canchipur - 795003, Manipur, India}

\emailAdd{$^{\dagger}$lkhumaba@gmail.com}
\emailAdd{$^\ast$yumnamprakash11@gmail.com}
\emailAdd{$^\ddagger$chandrachani@gmail.com}

\abstract{We examine whether or not the current observational preference for a dynamical dark energy(DE), as reported by DESI, is an artifact of the chosen functional form of dark energy equation of state (EoS) parameter $w(z)$ by considering a higher-order quadratic extension in EoS. We explore three distinct parameterizations of the DE models: the two-parameter (linear) Chevallier-Polarski-Linder (CPL) and Wang models, alongside a three-parameter parabolic (quadratic) formulation. Using a joint dataset of CMB measurements from Planck and ACT DR6, baryon acoustic oscillations of DESI DR2, and Type Ia supernovae from Pantheon+ and DES-SN5YR, we constrain the model parameters and observe a maximum frequentist preference of up to $\sim 4\sigma$ for these dynamical models. Specifically, the joint CMB + DESI + DES-SN5YR combination yields a substantial improvement in fit over $\Lambda\text{CDM}$, with $\Delta\chi^2_{\text{min}}$ values of $-20.5$, $-19.4$, and $-21.7$  for CPL, Wang, and Parabolic models, respectively. In addition, the Akaike Information Criterion (AIC) provides weak to no evidence for $\Lambda$CDM, consistently favoring the dynamical DE models instead. On the other hand, the Bayesian Information Criterion (BIC) imposes a heavy penalty on all the three models, reflecting its high sensitivity to dataset size and model complexity. We find at best a mild support for CPL and Wang models but the Parabolic model is strongly disfavored ($\Delta \text{BIC} \sim +6.4$ to $+15.6$). Furthermore, due to its expanded parameter space,  the Parabolic model exhibits a significant reduction in the Figure of Merit (FoM) compared to the two-parameter counterparts, demonstrating that the current cosmological datasets do not yet justify an introduction of the higher-order DE models. So, as a result, the two-parameter descriptions of $w(z)$ remain the optimal choice for testing the dynamical DE signatures.}

\begin{document}

\maketitle
\flushbottom

%+-+-+-+-+-+-+-+-+-+-+-+-+-+-+-+-+-+-+-+-+-+-+-+-+-+-

\input{Text/1_intro}

\input{Text/2_DynamicDarkEnergy}
\input{Text/3_datasetandMethodology}
\input{Text/4_results}
\input{Text/5_conclusions}

%+-+-+-+-+-+-+-+-+-+-+-+-+-+-+-+-+-+-+-+-+-+-+-+-+-+-

\acknowledgments
 LKM and YPS gratefully acknowledge Manipur University for the award of a Research Fellowship in support of their Ph.D. program. The results presented in this work were achieved using the HPC Cluster facilities at the Department of Physics, Manipur University and the HPC Pegasus Cluster of IUCAA, Pune. NCD acknowledges the Pegasus Cluster computer facility provided by IUCAA, Pune.

\appendix
\input{Text/6_appendix_w_f}

\bibliographystyle{JHEP}
\bibliography{references}

\end{document}

%% file: Text/1_intro.tex
% !TEX root = ../main.tex

 \section{Introduction}
\label{sec:intro}

In the late 1990s, two teams of astronomers discovered, through observations of distant Type Ia supernovae, that the expansion of the universe is accelerating rather than decelerating~\cite{Riess_1998, Perlmutter_1999}. The physical mechanism driving this late-time acceleration remains one of the most profound mysteries in modern cosmology. Various theoretical frameworks have been proposed under the umbrella term "dark energy", ranging from the simplest $\Lambda\text{CDM}$ model---characterized by a cosmological constant---to modifications of the Einstein field equations or an introduction of exotic scalar fields, all aimed at elucidating the nature of this cosmic phenomenon~\cite{Copeland}. Among these candidates, $\Lambda\text{CDM}$ has emerged as the most statistically favored model across a wide range of observations. Within the current cosmological paradigm, $\Lambda\text{CDM}$ is considered as the standard model. However, recent results from the Dark Energy Spectroscopic Instrument (DESI) collaboration have challenged this status quo, reporting a preference up to $4.2\sigma$ for a dynamical dark energy component over the constant vacuum energy ($\Lambda$) of the standard $\Lambda\text{CDM}$ model~\cite{tr6y-kpc6}. To investigate this potential evolution, the collaboration utilized the widely adopted two-parameter Chevallier-Polarski-Linder (CPL) parameterization~\cite{ChevallierPolarski2001, Linder2003}, which allows for a linear evolution of the equation of state (EoS) with respect to the scale factor. Since the release of the DESI data, a surge of studies has emerged utilizing these measurements to test a diverse array of EoS parameterizations—including the Barboza-Alcaniz (BA)~\cite{BarbozaAlcaniz}, Jassal-Bagla-Padmanabhan (JBP)~\cite{JassalBaglaPadmanabhan}, and exponential models~\cite{Dimakis, Pan}—as well as more exotic dark sector interactions and modified theories of gravity (e.g., Refs.~\cite{Giare_2024, lodha2025extendeddarkenergyanalysis}). From a more skeptical perspective, it has been argued that such a preference could be an artifact of a rigid functional form of EoS rather than a reflection of genuine cosmological dynamics~\cite{Wolf_2025}. This suggests that a prescribed parameterization may lack the necessary degrees of freedom to accurately capture the underlying physics, potentially leading to a spurious deviation from $\Lambda\text{CDM}$.

To further investigate and potentially consolidate the preference for a dynamical dark energy with the current data available, and also to test the viability of expanded EoS parameter spaces, we consider two additional dark energy parameterizations alongside CPL: the two-parameter Wang model~\cite{wang} and a three-parameter parabolic parameterization~\cite{Hu2014}. In this work, we first investigate these three dynamical dark energy parameterizations and place constraints on their parameters using a composite cosmological dataset: Baryon Acoustic Oscillation (BAO) data from the Dark Energy Spectroscopic Instrument (DESI)~\cite{DES:2024tys,DESI:2025zgx}, cosmic microwave background (CMB) measurements from the \textit{Planck} satellite~\cite{Aghanim:2019ame,rosenberg:2022} and the Atacama Cosmology Telescope (ACT)~\cite{Carron_2022,Qu_2024,Madhavacheril_2024}, and Type Ia supernovae measurements from the Pantheon+ compilation~\cite{Brout_2019,Brout:2022vxf} and the DES-SN5YR dataset~\cite{DES:2024tys}. Alongside, we also try to understand the constraining power of different dataset particularly with the combinations of: 1) CMB + DESI, 2) CMB + DESI + Pantheon+, and 3) CMB + DESI + DES-SN5YR.

We perform a comparative analysis of these models—measuring their performance relative to the $\Lambda\text{CDM}$ baseline—to evaluate their relative efficacy in capturing current cosmological observations and determine if the reported dynamical nature of dark energy is robust across different functional forms. To assess model performance, we calculate the $\chi^2$ statistic, N$\sigma$ statistical significance, figure of merit~\cite{wang}. Furthermore, we utilize the Akaike information criterion (AIC)~\cite{Akaike:1974} and the Bayesian information criterion (BIC)~\cite{BIC:schwarz1978estimating} to rigorously evaluate model selection and penalize over-parameterization for the mentioned three dataset combinations.

The paper is organized as follows. We introduce the parametric forms of the dynamical dark energy models in section~\ref{sec:dde}. Section~\ref{sec:dataMethods} describes the dataset and methodology used for our analysis. In section~\ref{sec:results}, we present our results and compare them with respect to the $\Lambda$CDM model.  Finally, we summarize and conclude in section~\ref{sec:conclusion}. For completeness, the reconstructed posterior distributions for both the dark energy EoS and its energy density are provided in the Appendix~\ref{sec:w_f}.

%% file: Text/2_DynamicDarkEnergy.tex
% !TEX root = ../main.tex

 \section{Parameterizations of Dynamic Dark Energy}
\label{sec:dde}

Considering a spatially flat homogeneous and isotropic universe described by the Friedmann-Lemaître-Robertson-Walker (FLRW) spacetime metric, the expansion of our universe governed by Einstein's field equations are as follows:
% ---------------------
\begin{equation}
H^{2} =\frac{\dot a^2(t)}{a^2(t)}= \frac{8\pi G}{3} \rho_{t},
\label{eq1}
\end{equation}
% ---------------------
\begin{equation}
\dot H = - {4\pi G} (\rho_{t} + P_t),
\end{equation}
% ---------------------
where $a(t)$ is the dimensionless scale factor that describes the expansion of the universe.  $\rho_t$ and $P_t$ denote the total energy density and the total pressure of the universe, respectively. Mainly, our universe consists of radiation $\rho_r$, baryonic matter $\rho_b$, cold dark matter $\rho_c$ and dark energy $\rho_{DE}$, so $\rho_t = \rho_b +\rho_c +\rho_r+\rho_{DE}$. Assuming that the dark energy behaves as a perfect fluid, its physical properties is characterized by the equation of state (EoS) parameter, $w$ as:
% ---------------------
\begin{equation}
w = \frac{P_{DE}}{\rho_{DE}},
\end{equation}
% ---------------------
where $P_{DE}$ represents the pressure of the dark energy fluid. This parameter, $w$, dictates the dynamics of the dark energy field. From Eq.~\eqref{eq1}, the dimensionless expansion rate, $E(a)$, can be written as:
% ---------------------
\begin{equation}
\label{expansion}
E^{2}(a) \equiv \frac{H^{2}(a)}{H_{0}^{2}} = \Omega_{m0} a^{-3} + \Omega_{r0} a^{-4} + \Omega_{k0} a^{-2} + \Omega_{\text{DE,0}} F(a),
\end{equation}
% ---------------------
where $\Omega_{m0} = (\Omega_{b0}+\Omega_{c 0})$, $\Omega_{r0}$, $\Omega_{k0}$, and $\Omega_{\text{DE,0}}$ represent the current dimensionless density parameters for matter including baryonic matter and cold dark matter, radiation, spatial curvature, and dark energy, respectively. The function $F(a)$ characterizes the evolution of the dark energy density relative to its value today:
% ---------------------
\begin{equation}
\label{eq:f}
F(a) \equiv \dfrac{\rho_{\rm DE}(a)}{\rho_{\rm DE,0}}
= \exp\!\left[
3 \int_{a}^{1} \dfrac{1 + w(a')}{a'} \, da'
\right].
\end{equation}
%------------------------------------------

While the $\Lambda\text{CDM}$ model with $w(a) = -1$ and $F(a) = 1$ serves as the standard cosmological concordance model, parameterizing the cosmic evolution with a minimal set of parameters, it faces immense pressure from multiple internal and external inconsistencies. Currently, it is challenged by empirical discrepancies, such as the Hubble ($H_{0}$) tension and the $\sigma_{8}$ clustering discrepancy, and theoretical anomalies such as the fine-tuning of the cosmological constant value and the cosmic coincidence problem.

In our work, we employ three dark energy parameterizations to rigorously test for the dynamical nature of dark energy field. Firstly, we consider the widely adopted Chevallier-Polarski-Linder (CPL) parameterization~\cite{ChevallierPolarski2001, Linder2003}, which is characterized by the Equation of State (EoS) as:
% ---------------------
\begin{equation}
w(a) = w_{0} + w_{a} (1 - a),
\end{equation}
% ---------------------
where $w_{0}$ is the current EoS value and $w_{a}$ represents the rate of its evolution; a non-zero $w_{a}$ signifies dynamical dark energy. This model recovers the standard $\Lambda$CDM paradigm in the limit $(w_{0}, w_{a}) = (-1, 0)$. Given its central role in the recent DESI collaboration~\cite{tr6y-kpc6} reports regarding dynamical dark energy, we reproduce the similar results here to compare our findings within the current cosmological context. Furthermore, including the CPL model allows us to check a rigorous performance comparison against the other 2-parameters and 3-parameters models considered in this work, enabling to evaluate which mathematical form captures best the expansion history suggested by the DESI DR2 dataset.

The second parameterization we used is a two-parameters model introduced by Wang~\cite{wang}, hereafter referred to as the Wang model. This formulation was developed to characterize the Equation of State (EoS) using a set of minimally correlated parameters, thereby reducing the degeneracies often encountered in the CPL framework. The EoS in the Wang model is expressed as:
% ---------------------
\begin{equation}
\label{eqn:wang}
w(a) = 3 w_{a} -2 w_{0} + 3(w_{0}-w_{a}) a,
\end{equation}
% ---------------------
where $w_{0}$ again denotes the EoS value at the present day ($a=1$) and $w_{a}$ represents the EoS value at redshift $z=0.5$ (corresponding to $a=2/3$). In this formulation, the standard $\Lambda\text{CDM}$ paradigm is recovered in the limit of $(w_{0}, w_{a}) = (-1, -1)$. We include this model to evaluate whether a different linear structure in the scale factor $a$ affects the statistical significance of the dynamical dark energy preference.

The final model under consideration is a three-parameters parabolic formulation, introduced by Hu et al.~\cite{Hu2014} to investigate whether observational data favor a non-monotonic or ``turning'' behavior in the dark energy equation of state. 
The EoS for this parabolic model is given by:
% ---------------------
\begin{equation}
w(a) = w_{0} + w_{a} (a_{t}-a)^{2},
\end{equation}
% ---------------------
where $a_{t}$ specifies the scale factor at which the turning point occurs, $w_{0}$ represents the value of the EoS at that extremum ($w(a_t) = w_0$) and $w_{a}$ determines whether the parabola opens upward or downward. The standard $\Lambda\text{CDM}$ paradigm is recovered in the limit $(w_{0}, w_{a}) = (-1, 0)$. By including this three-parameter model, we facilitate a comparative study between simpler two-parameter EoS forms and more complex descriptions. This allows us to evaluate whether the current data justify the inclusion of additional degrees of freedom to capture the evolution of dark energy.

%% file: Text/3_datasetandMethodology.tex
% !TEX root = ../main.tex

 \section{Datasets and Methodology}
\label{sec:dataMethods}

The datasets used in our analysis are the Cosmic Microwave Background Radiation(CMB), the baryon acoustic oscillation(BAO) data and the supernovae (SNIa) observations:
%%+-+-+-+-+-+-+-+-+-+-+-+-+-+-+-+-+-+-
\begin{itemize}

\item \textit{Cosmic Microwave Background Radiation(CMB):} 

This dataset consists of CMB temperature, polarization and lensing power spectra from the \textit{Planck} satellite and an additional small scales lensing measurements from the Atacama Cosmology Telescope (ACT) Data Release 6 (DR6). Specifically, in the low-multipole ($2 \leq \ell < 30$) regime, we utilize temperature ($TT$) likelihoods called \texttt{Commander} and $E$-mode ($EE$) polarization likelihoods from \texttt{SimAll}~\cite{Aghanim:2019ame}. And in the high-$\ell$ regime ($30 \leq \ell \leq 2500$), we employ the high-$\ell$ \texttt{CamSpec} ($TTTEEE$) likelihood based on the \textit{Planck} \texttt{NPIPE} release~\cite{rosenberg:2022}. These are complemented by the ACT DR6 Lensing Likelihood \footnote{v1.2: \url{https://github.com/ACTCollaboration/act_dr6_lenslike}} using the \texttt{actplanck\_baseline} variant, which incorporates both ACT and \textit{Planck} lensing data~\cite{Carron_2022,Qu_2024,Madhavacheril_2024}.

\item \textit{Baryon acoustic oscillations(BAO):}  

We utilize the Baryon Acoustic Oscillation (BAO) measurements from the Dark Energy Spectroscopic Instrument (DESI) Data Release 2 (DR2). This dataset, hereafter referred to as DESI, comprises results from the first three years of the survey, spanning a redshift range of $0.1 < z < 3.5$. It includes the angular diameter distance measurements derived from the Bright Galaxy Survey (BGS), Luminous Red Galaxies (LRG), Emission Line Galaxies (ELG),quasars (QSOs), and the Lyman-$\alpha$ forest~\cite{tr6y-kpc6}.

\item \textit{Type 1a Supernovae (SNe~Ia):}
 SNe Ia being standard candles measures the distance modulus at a particular redshift and determines the expansion history of the universe. For our work, we consider two different sets of compilation:

%+-+-+-+-+-+-+-+-+-+-+-+-+-+-+-+-+-+-
\begin{itemize}

\item The \textit{Pantheon+} compilation~\cite{Brout:2022vxf} consists of light curves from 1550 distinct, spectroscopically classified SNe~Ia drawn from a variety of observational surveys, including those from the first three years of the Dark Energy Survey (DES)~\cite{Brout_2019}. The complete collection spans a broad redshift range of $0.001 < z < 2.26$.

\item The \textit{DES-SN5YR} dataset contains 1829 SNe~Ia, 1635 of which are drawn from the full five years of the Dark Energy Survey (DES) Supernova Program~\cite{DES:2024tys}. These events are photometrically-classified and span a redshift range of $0.10 < z < 1.13$. The rest of the sample consists of low-redshift ($0.025 < z < 0.10$) SNe~Ia which are sourced from external surveys.

\end{itemize}
%%+-+-+-+-+-+-+-+-+-+-+-+-+-+-+-+-+-+-

\end{itemize}
%%+-+-+-+-+-+-+-+-+-+-+-+-+-+-+-+-+-+-
Abdul-Karim et al.~\cite{tr6y-kpc6} demonstrated utilizing the CPL parameterization that the DESI dataset alone provides only a weak preference for the dynamical nature of dark energy. However, the inclusion of the DES-SN5YR compilation in combination with CMB and DESI BAO results yields a substantial $4.2\sigma$ preference for dynamical dark energy. In order to understand the constraining power of different datasets, we constrain the models using three different data combinations: (i) CMB + DESI, (ii) CMB + DESI + Pantheon+, and (iii) CMB + DESI + DES-SN5YR.

First, we implement the aforementioned dark energy parameterizations within the \texttt{CAMB} Boltzmann solver~\cite{Lewis:1999bs,Howlett:2012mh}\footnote{\url{https://github.com/cmbant/camb}}, which serves as our primary theory code. For parameter estimation across these models, including $\Lambda \text{CDM}$, we perform Markov Chain Monte Carlo (MCMC) sampling via the \texttt{Cobaya} framework~\cite{Torrado:2020dgo}\footnote{\url{https://github.com/CobayaSampler/cobaya}}, employing the fast-dragging technique~\cite{Neal:2005}. Throughout the analysis, we assume a flat universe ($\Omega_k = 0$) and adopt flat priors for all sampled parameters as summarised in Table \ref{tab:priors}. The convergence of the chains is rigorously evaluated using the Gelman-Rubin diagnostic~\cite{GelmanRubin}, ensuring that $R - 1 < 0.01$. Subsequently, the resulting posterior distributions and parameter constraints are analyzed using the \texttt{GetDist} package~\cite{Lewis_2025}.

In addition, we check the relative performance of the CPL, Wang, and Parabolic models with respect to  the $\Lambda\text{CDM}$ model by comparing $\Delta \chi^2$ where $\Delta\chi^2$ is the difference between the best-fit $\chi^2$ of the model and that of $\Lambda\text{CDM}$: $\Delta\chi^2 = \chi^2_{\text{model,i}} - \chi^2_{\Lambda\text{CDM}}.$ A more negative value of $\Delta\chi^2$ means a lower value $ \chi^2_{\text{model,i}}$ compared to $\chi^2_{\Lambda\text{CDM}}$, which represents a better fit to the data we used. This $\Delta\chi^2$ is further expressed in terms of an $N\sigma$ significance value for a 1D Gaussian distribution via the Cumulative Distribution Function (CDF) as:
% ----------------------
\begin{equation}
P(\Delta\chi^2, \nu) = \text{erf}\left( \frac{N}{\sqrt{2}} \right),
\end{equation}
% ----------------------
where $P$ is the $\chi^2$ CDF for $\nu$ degrees of freedom which is given by the number of extra parameters each model possesses relative to the $\Lambda\text{CDM}$ baseline.

To perform model selection and penalize over-parameterization, we employ  Akaike information criterion (AIC)~\cite{Akaike:1974} and the Bayesian information criterion (BIC)~\cite{BIC:schwarz1978estimating}. Both metrics are established benchmarks for cosmological model selection (see Refs.~\cite{Liddle:2004,Liddle:2007,Arevalo2017,Forconietal2025}), where a lower value of AIC and BIC indicates a superior balance between goodness-of-fit and parametric complexity, signifying a preferred model for the given observational data. The AIC is defined as:
% ----------------------
\begin{equation}
\text{AIC} = \chi^2_{\min} + 2 k,
\end{equation}
% ----------------------
 where $\chi^2_{\min}$ is the minimum (best-fit) $\chi^2$ value of the model and $k$ is the number of free parameters in the model. Following the evidence threshold ranges adopted in Ref.~\cite{Arevalo2017}, we evaluate the differences in Information Criteria ($\Delta\text{IC}$) relative to the $\Lambda\text{CDM}$ model. We define $\Delta \text{AIC} = \text{AIC}_{\Lambda\text{CDM}} - \text{AIC}_{\text{model,i}}$ and interpret the values as follows: $0 \leq \Delta \text{AIC} < 2$ indicates "strong evidence" in favor of the $\Lambda$CDM model,  i.e "weak evidence" in favor of the model i, $4 < \Delta \text{AIC} \leq 7$ indicates "weak evidence" in favor of the $\Lambda$CDM model, correspondingly "positive evidence" in favor of the  model i, and $\Delta \text{AIC} > 10$ indicates "no evidence" in favor of the $\Lambda$CDM model and "strong evidence" in favor of the new model i. On the other hand, the BIC imposes a stricter penalty based on the dataset size as it is defined by
 % ----------------------
\begin{equation}
\text{BIC} = \chi^2_{\min} + k \ln(N),
\end{equation}
% ----------------------
where $N$ is the total number of data points used in the joint likelihood. For BIC, we define $\Delta \text{BIC} = \text{BIC}_{\text{model,i}} - \text{BIC}_{\Lambda\text{CDM}}$ with the thresholds interpreted as "evidence against" the model i compared to the ${\Lambda\text{CDM}}$ model as follows: $0 \leq \Delta \text{BIC} < 2$ indicates "negligible evidence against" , $2 \leq \Delta \text{BIC} < 6$ indicates "positive evidence against", $6 \leq \Delta \text{BIC} < 10$ indicates "strong evidence against", and  $\Delta \text{BIC} > 10$ indicates "very strong evidence against" the model i with respect to the ${\Lambda\text{CDM}}$ model.

We further evaluate the constraining power of our datasets using a relative generalized Figure of Merit (FoM) as proposed in Ref.~\cite{wang}:
% ----------------------
\begin{equation*}
\text{FoM} = \frac{1}{\sqrt{\text{det}(\mathbf{S})}},
\end{equation*}
% ----------------------
where $\mathbf{S}$ represents the covariance matrix of the dark energy parameters (specifically $w_0$, $w_a$, and $a_t$). This generalized formulation is particularly advantageous as it accounts for non-Gaussian behavior in the posterior distributions, providing a more robust measure of the constraining power of the CMB, BAO and supernova datasets across different parameterizations and dataset combinations.

% ----------------------
\begin{table}[tbp]
    \centering
    \setlength{\tabcolsep}{30pt}      % Increases column width
    \begin{tabular}{|l|c|}
        \hline
        \rule{0pt}{3ex}
%        \\[0.5ex]
\textbf{Parameters}      & \textbf{Prior}\\
        \hline
        \rule{0pt}{3ex}
$\log(10^{10} A_\mathrm{s})$     & $[1.61, 3.91]   $ \\ 
$n_\mathrm{s}$                   & $[0.8, 1.2]     $ \\
$100\theta_\mathrm{MC}$          & $[0.5, 10]      $ \\
$\Omega_\mathrm{b} h^2$          & $[0.005, 0.1]   $ \\ 
$\Omega_\mathrm{c} h^2$          & $[0.001, 0.99]  $ \\
$\tau_\mathrm{reio}$             & $[0.01, 0.8]    $ \\
\hline
$w_0$ (CPL)                      & $[-3, 1]        $ \\
$w_a$ (CPL)                      & $[-3, 2]        $ \\ 
\hline
$w_0$ (Wang)                     & $[-1.5, -0.5]   $ \\
$w_a$ (Wang)                     & $[-1.2, -0.7]   $ \\ 
\hline
$w_0$ (Parabola)                 & $[-2,0]         $ \\
$w_a$ (Parabola)                 & $[-10, 5]       $ \\ 
$a_{t}$ (Parabola)               & $[-2,4]         $ \\ 
        \hline
    \end{tabular} 
\caption{Prior for the cosmological parameters and model parameters used in the MCMC analysis. All distributions are assumed to be uniform (flat) within the specified boundaries.}
\label{tab:priors}
\end{table}

% ----------------------

%% file: Text/4_results.tex
% !TEX root = ../main.tex

 \section{Results}
\label{sec:results}
%\subsection{CPL}
In this section, we present the cosmological constraints obtained from our MCMC analysis, which was executed using a publicly available code called \texttt{Cobaya}~\cite{Torrado:2020dgo} and analyzed via the \texttt{GetDist} software package \cite{Lewis_2025}. We begin by examining the Chevallier-Polarski-Linder (CPL) parameterization, followed by the Wang and parabolic models. Finally, we provide a comparative overview of these models in Table~\ref{tab:comparision}, utilizing the various statistical metrics discussed in section~\ref{sec:dataMethods}. The priors
used for our analysis are presented in Table.\ref{tab:priors}.

\begin{figure}[tbp]
    \centering
    \includegraphics[width=\textwidth]{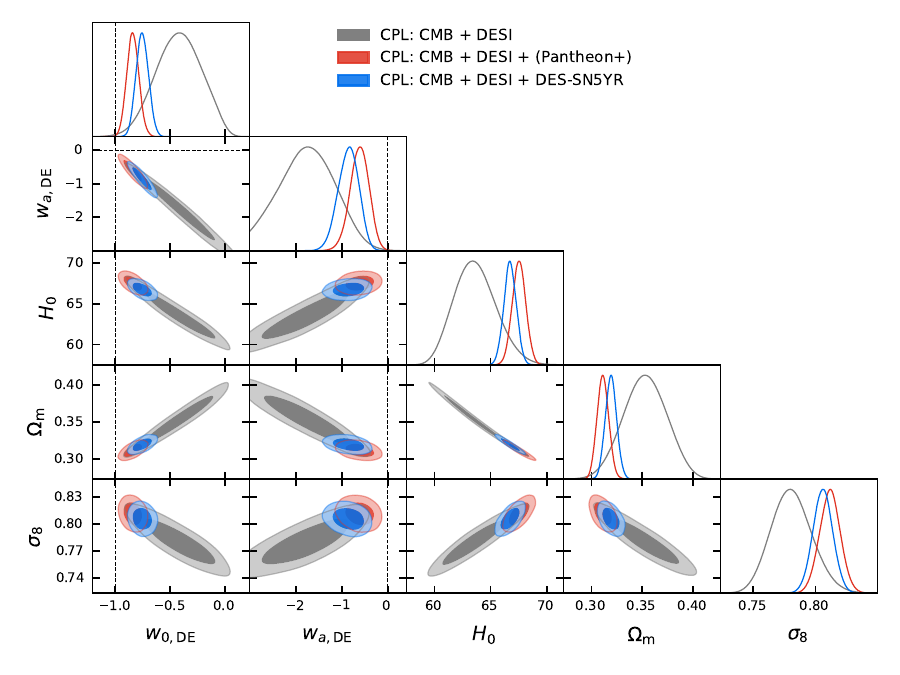}
    \caption{Triangle plot depicting 68\% and 95\% confidence level(CL) for selected cosmological parameters along with the model parameters alongside their 1D marginalized posterior distributions under the CPL parameterization. Contours are shown for three dataset combinations: 1) CMB + DESI (gray), 2) CMB + DESI + Pantheon+ (red), and 3) CMB + DESI + DES-SN5YR (blue). The black dashed lines indicate the baseline $\Lambda\text{CDM}$ values, $(w_{0}, w_{a}) = (-1, 0)$.}
    \label{fig:CPL}
\end{figure}

Under the CPL parameterization, the marginalized posterior constraints for the key cosmological parameters and model parameters are summarized in Table~\ref{tab:cpl} at 68\% confidence level (CL) for the three dataset combinations mentioned above. The associated 1D and 2D posterior distributions (68\% and 95\% CL) are shown in the triangular plot provided in Figure~\ref{fig:CPL}. As expected, we recover the findings of the DESI Collaboration, see figure 11 and also equations (25), (26) and (28) of Ref.~\cite{tr6y-kpc6} to compare the constraints values of $(w_0, w_a)$ provided by CMB+DESI, CMB + DESI CMB + DESI +(Pantheon+) and CMB + DESI + DES-SN5YR respectively. Here we present the constraints in full detail to facilitate a direct comparison with the other alternative dark energy models. The black dashed lines in Figure~\ref{fig:CPL} represent the baseline $\Lambda\text{CDM}$ values, $(w_{0}, w_{a}) = (-1, 0)$. The $\Lambda\text{CDM}$ model, located at the intersection of these lines, lies well outside the $2\sigma$ confidence region across all considered dataset combinations. This departure is particularly significant for the joint CMB + DESI + DES-SN5YR data, which exhibits the largest deviation from $\Lambda\text{CDM}$, thereby favoring the dynamical nature of dark energy up to 4$\sigma$ preference (see Table~\ref{tab:comparision}).
%------------TABLE---------------
\begin{table}[tbp]
    \centering
    \begin{tabular}{|c|c|c|c|}
        \hline
        \rule{0pt}{3ex}
        \multirow{2}{*} {\bf Parameter}
        &\multirow{2}{*}{\bf CMB+DESI} 
        &\multirow{2}{*}[0.8ex]{\bf CMB+DESI}
        &\multirow{2}{*}[0.8ex]{\bf CMB+DESI} 
        \\
        &
        &\bf +(Pantheon+)
        & \bf +DES-SN5YR
        \\[0.5ex]
        \hline
        \rule{0pt}{3ex}

{$\log(10^{10} A_\mathrm{s})$}    & $3.038\pm 0.013            $      & $3.043\pm 0.013            $    & $3.042\pm 0.013            $\\
{$n_\mathrm{s}   $}               & $0.9646\pm 0.0037          $      & $0.9659\pm 0.0036          $    & $0.9654\pm 0.0036          $\\
{$100\theta_\mathrm{MC}$}         & $1.04080\pm 0.00024        $      & $1.04086\pm 0.00024        $    & $1.04084\pm 0.00024        $\\
{$\Omega_\mathrm{b} h^2$}         & $0.02221\pm 0.00012        $      & $0.02225\pm 0.00013        $    & $0.02223\pm 0.00013        $\\
{$\Omega_\mathrm{c} h^2$}         & $0.11959\pm 0.00083        $      & $0.11904\pm 0.00081        $    & $0.11923\pm 0.00081        $\\
{$w_{0,\mathrm{DE}}$}             & $-0.42\pm 0.20             $      & $-0.839\pm 0.055           $    & $-0.753\pm 0.056           $\\
{$w_{a,\mathrm{DE}}$}             & $-1.74\pm 0.58             $      & $-0.62^{+0.21}_{-0.19}     $    & $-0.86^{+0.24}_{-0.21}     $\\
{$\tau_\mathrm{reio}$}            & $0.0526\pm 0.0070          $      & $0.0551\pm 0.0071          $    & $0.0542\pm 0.0072          $\\
$H_0                       $      & $63.6^{+1.7}_{-2.0}        $      & $67.53\pm 0.60             $    & $66.73\pm 0.55             $\\
$\Omega_\mathrm{m}         $      & $0.353\pm 0.021            $      & $0.3113\pm 0.0057          $    & $0.3192\pm 0.0055          $\\
$\sigma_8                  $      & $0.781^{+0.015}_{-0.017}   $      & $0.8117\pm 0.0083          $    & $0.8059\pm 0.0080          $\\
${\rm{Age}}/\mathrm{Gyr}   $      & $13.767\pm 0.020           $      & $13.760\pm 0.020           $    & $13.760\pm 0.020           $\\
$r_\mathrm{drag}           $      & $147.39\pm 0.21            $      & $147.49\pm 0.21            $    & $147.46\pm 0.21            $\\        
        
        \hline
    \end{tabular}
\caption{Marginalized constraints at 68\% confidence level (CL) for the sampled and derived cosmological parameters within the CPL parameterization using three dataset combinations:1) CMB + DESI, 2) CMB + DESI + Pantheon+, and 3) CMB + DESI + DES-SN5YR.
}
\label{tab:cpl}
\end{table}
%-------------------------------------
%-------------------------------------
We observe that combining the DESI BAO (DR2) measurements with the CMB data alone provides a loose bound on almost all cosmological parameters (see the gray contour in Figure~\ref{fig:CPL}) and yields the following marginalized posterior constraints for the EoS parameters:
% ---------------------------
\begin{subequations}\label{eq:eos_cpl_1}
\begin{align}
\label{eq:eos_cpl_1:w0}
w_{0} & = -0.42 \pm 0.20 
\\
\label{eq:eos_cpl_1:wa}
w_{a} & = -1.74 \pm 0.58.
\end{align}
\end{subequations}
% ---------------------------
However, the inclusion of Type Ia supernova data in the joint CMB + DESI BAO analysis significantly tightens the constraints (see the red and blue contours in Figure~\ref{fig:CPL}).  Specifically, incorporating the Pantheon+ sample results in a shift of the marginalized EoS parameters, $w_0$ and $w_a$ by 99\% and 64\%, respectively, compared to the results obtained without supernova constraints:
% ---------------------------
\begin{subequations}\label{eq:eos_cpl_2}
\begin{align}
\label{eq:eos_cpl_2:w0}
w_{0} & = -0.839\pm 0.055
\\
\label{eq:eos_cpl_2:wa}
w_{a} & = -0.62^{+0.21}_{-0.19}.
\end{align}
\end{subequations}
% ---------------------------
Similarly, incorporating the DES-SN5YR supernova dataset (replacing Pantheon+) results in a less pronounced shift, with the marginalized $(w_0, w_a)$ parameters moving by 79\% and 50\%, respectively, relative to the CMB + DESI BAO baseline values:
% ---------------------------
\begin{subequations}\label{eq:eos_cpl_3}
\begin{align}
\label{eq:eos_cpl_3:w0}
w_{0} & = -0.753\pm 0.056
\\
\label{eq:eos_cpl_3:wa}
w_{a} & = -0.86^{+0.24}_{-0.21}.
\end{align}
\end{subequations}
% ---------------------------

The supernova data also influences the constraints on the present expansion rate, $H_0$ and the growth of structure. Particularly, the $H_0$ value increases by 6\% when the Pantheon+ dataset is added to the CMB + DESI combination, shifting from a baseline value of $63.6$ km s$^{-1}$ Mpc$^{-1}$ to $67.53$ km s$^{-1}$ Mpc$^{-1}$. Again, the addition of DES-SN5YR results in increase of 5\% in the  $H_0$ value. A similar trend is observed for the amplitude of matter fluctuations, $\sigma_8$. Its marginalized posterior value increases by 4\%  and 3\% with the inclusion of Pantheon+ and DES-SN5YR alongside the CMB + DESI baseline respectively.

%----------------------------

\begin{table}[tbp]
    \centering
    \begin{tabular}{|c|c|c|c|}
        \hline
        \rule{0pt}{3ex}
        \multirow{2}{*} {\bf Parameter}
        &\multirow{2}{*}{\bf CMB+DESI} 
        &\multirow{2}{*}[0.8ex]{\bf CMB+DESI}
        &\multirow{2}{*}[0.8ex]{\bf CMB+DESI} 
        \\
        &
        &\bf +(Pantheon+)
        & \bf +DES-SN5YR
        \\[0.5ex]
        \hline
        \rule{0pt}{3ex}

{$\log(10^{10} A_\mathrm{s})$}     & $3.041\pm 0.013            $    & $3.044\pm 0.013            $    & $3.042\pm 0.013            $\\
{$n_\mathrm{s}   $}                & $0.9653\pm 0.0035          $    & $0.9659\pm 0.0037          $    & $0.9655\pm 0.0037          $\\
{$100\theta_\mathrm{MC}$}          & $1.04084\pm 0.00024        $    & $1.04087\pm 0.00024        $    & $1.04085\pm 0.00024        $\\
{$\Omega_\mathrm{b} h^2$}          & $0.02223\pm 0.00012        $    & $0.02225\pm 0.00012        $    & $0.02223\pm 0.00012        $\\
{$\Omega_\mathrm{c} h^2$}          & $0.11927\pm 0.00080        $    & $0.11904\pm 0.00081        $    & $0.11923\pm 0.00083        $\\
{$w_{0,\mathrm{DE}}$}              & $> -0.671                  $    & $-0.839\pm 0.054           $    & $-0.752\pm 0.057           $\\
{$w_{a,\mathrm{DE}}$}              & $-1.017^{+0.042}_{-0.037}  $    & $-1.044^{+0.033}_{-0.030}  $    & $-1.039^{+0.035}_{-0.031}  $\\
{$\tau_\mathrm{reio}$}             & $0.0541\pm 0.0071          $    & $0.0552\pm 0.0071          $    & $0.0543\pm 0.0071          $\\
$H_0                       $       & $65.53^{+0.75}_{-1.4}      $    & $67.53\pm 0.59             $    & $66.75\pm 0.55             $\\
$\Omega_\mathrm{m}         $       & $0.331^{+0.014}_{-0.0080}  $    & $0.3113\pm 0.0056          $    & $0.3191\pm 0.0055          $\\
$\sigma_8                  $       & $0.796^{+0.010}_{-0.013}   $    & $0.8117\pm 0.0083          $    & $0.8060\pm 0.0079          $\\
${\rm{Age}}/\mathrm{Gyr}   $       & $13.764\pm 0.020           $    & $13.760\pm 0.019           $    & $13.759\pm 0.019           $\\
$r_\mathrm{drag}           $       & $147.45\pm 0.20            $    & $147.49\pm 0.21            $    & $147.46\pm 0.21            $\\

        \hline
    \end{tabular}
\caption{Marginalized constraints at 68\% CL for the sampled and derived cosmological parameters within the Wang parameterization using three dataset combinations:1) CMB + DESI, 2) CMB + DESI + Pantheon+, and 3) CMB + DESI + DES-SN5YR.
}
\label{tab:wang}
\end{table}

\begin{figure}[tbp]
    \centering
    \includegraphics[width=\textwidth]{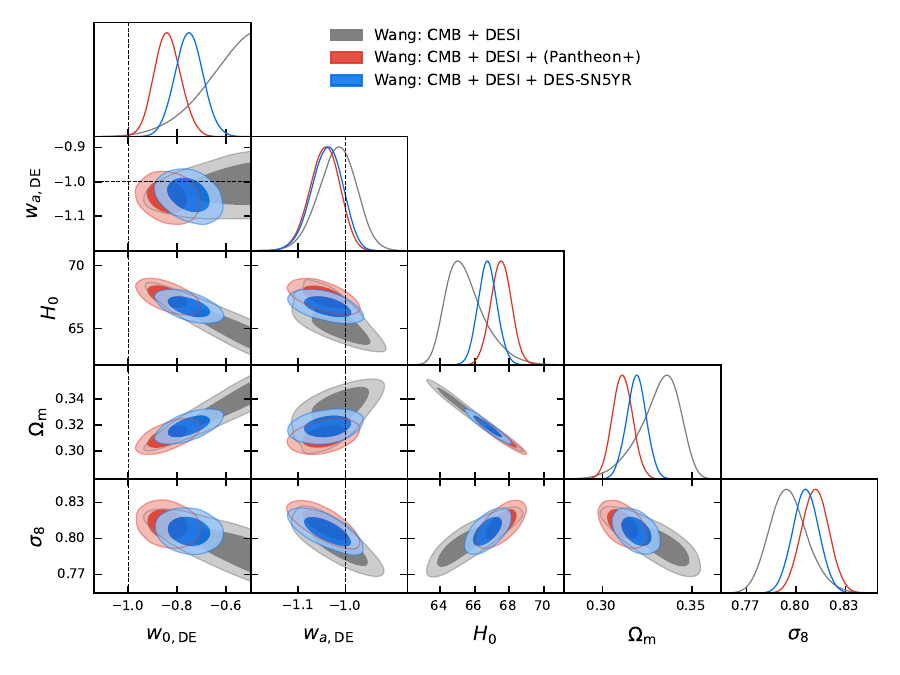}
    \caption{Triangle plot depicting 68\% and 95\% CL of the cosmological parameters along with 1D marginalised posterior distribution for Wang parameterization using three combinations of datasets: CMB + DESI (gray), CMB + DESI + Pantheon+ (red), and CMB + DESI + DES-SN5YR (blue). The black dashed lines indicate the baseline $\Lambda\text{CDM}$ values, $(w_{0}, w_{a}) = (-1, -1)$.}
    \label{fig:wang}
\end{figure}

%----------------------------
In Table~\ref{tab:wang}, we summarize the marginalized posterior constraints for key cosmological parameters and model parameters under the Wang parameterization at 68\% CL for the three dataset combinations. The associated 1D and 2D posterior distributions are provided in the Figure~\ref{fig:wang} where the black dashed lines indicate the baseline $\Lambda\text{CDM}$ values, $(w_{0}, w_{a}) = (-1, -1)$. As observed in the CPL model, the baseline $\Lambda\text{CDM}$ point in the Wang model as shown in Figure~\ref{fig:wang} also lies well outside the $2\sigma$ confidence region across all dataset combinations. This tension is most pronounced in the joint CMB + DESI + DES-SN5YR analysis (see Table~\ref{tab:comparision}), further reinforcing the preference for the dynamical dark energy framework upto $4\sigma$ over a constant $\Lambda$.Similar to CPL results, a combination of DESI BAO (DR2) measurements with CMB data alone yields poor constraints on the Equation of State (EoS) parameters  (grey contours in Figure~\ref{fig:wang}):
% ---------------------------
\begin{subequations}\label{eq:eos_wang_1}
\begin{align}
\label{eq:eos_wang_1:w0}
w_{0} & > -0.671 
\\
\label{eq:eos_wang_1:wa}
w_{a} & = -1.017^{+0.042}_{-0.037}.
\end{align}
\end{subequations}
% ---------------------------
Again, the inclusion of Type Ia supernova data to the CMB + DESI BAO analysis constrains $w_0$ effectively, leading to a robust constraint on the EoS parameters (red and blue contours in the figure \ref{fig:wang}). The Pantheon+ sample results in a minor shift in $w_a$ parameter---approximately 2.6\%---relative to the CMB + DESI BAO baseline:
% ---------------------------
\begin{subequations}\label{eq:eos_wang_2}
\begin{align}
\label{eq:eos_wang_2:w0}
w_{0} & = -0.839\pm 0.054
\\
\label{eq:eos_wang_2:wa}
w_{a} & = -1.044^{+0.033}_{-0.030}
\end{align}
\end{subequations}
% ---------------------------
and the DES-SN5YR supernova dataset results in an even smaller shift of $w_a$ parameter by 2.2\% relative to the CMB + DESI BAO baseline:
% ---------------------------
\begin{subequations}\label{eq:eos_wang_3}
\begin{align}
\label{eq:eos_wang_3:w0}
w_{0} & = -0.752\pm 0.057
\\
\label{eq:eos_wang_3:wa}
w_{a} & = -1.039^{+0.035}_{-0.031}.
\end{align}
\end{subequations}
% ---------------------------

Previous constraints on the Wang parameterization primarily utilized BAO measurements from BOSS DR12~\cite{SixiangWen_2018} and the analysis demonstrated a significantly higher Figure of Merit (FoM) compared to the CPL model, the data at the time remained consistent with the $\Lambda$CDM paradigm, offering negligible evidence for a dynamical dark energy component. In our analysis, however, the inclusion of BAO measurements from DESI DR2 data yields arguably compelling evidence for the dynamical nature of dark energy, marking a significant departure from the standard cosmological model.

% ---------------------------

%===========================================

\begin{table}[tbp]
    \centering
    \begin{tabular}{|c|c|c|c|}
        \hline
        \rule{0pt}{3ex}
        \multirow{2}{*} {\bf Parameter}
        &\multirow{2}{*}{\bf CMB+DESI} 
        &\multirow{2}{*}[0.8ex]{\bf CMB+DESI}
        &\multirow{2}{*}[0.8ex]{\bf CMB+DESI} 
        \\
        &
        &\bf +(Pantheon+)
        & \bf +DES-SN5YR
        \\[0.5ex]
        \hline
        \rule{0pt}{3ex}

{$\log(10^{10} A_\mathrm{s})$}                   & $3.039\pm 0.013            $ & $3.040\pm 0.013            $   & $3.042\pm 0.013            $\\
{$n_\mathrm{s}   $}                              & $0.9648\pm 0.0037          $ & $0.9652\pm 0.0037          $   & $0.9655\pm 0.0036          $\\
{$100\theta_\mathrm{MC}$}                        & $1.04080\pm 0.00024        $ & $1.04083\pm 0.00024        $   & $1.04084\pm 0.00024        $\\
{$\Omega_\mathrm{b} h^2$}                        & $0.02221\pm 0.00013        $ & $0.02223\pm 0.00013        $   & $0.02224\pm 0.00012        $\\
{$\Omega_\mathrm{c} h^2$}                        & $0.11951\pm 0.00088        $ & $0.11935\pm 0.00087        $   & $0.11922\pm 0.00084        $\\
{$w_{0,\mathrm{DE}}$}                            & $< -0.666                  $ & $-0.889^{+0.075}_{-0.068}  $   & $-0.99^{+0.43}_{-0.52}     $\\
{$w_{a,\mathrm{DE}}$}                            & $-1.9^{+3.4}_{-1.1}        $ & $-2.1^{+2.5}_{-1.1}        $   & $-0.52^{+1.4}_{-0.54}      $\\
{$a_{t}          $}                              & $0.78^{+0.65}_{-0.93}      $ & $0.93\pm 0.70              $   & $0.7^{+1.1}_{-1.2}         $\\
{$\tau_\mathrm{reio}$}                           & $0.0530\pm 0.0072          $ & $0.0536\pm 0.0072          $   & $0.0543\pm 0.0071          $\\
$H_0                       $                     & $65.7^{+1.9}_{-2.7}        $ & $67.66\pm 0.61             $   & $66.82\pm 0.57             $\\
$\Omega_\mathrm{m}         $                     & $0.331^{+0.025}_{-0.022}   $ & $0.3107\pm 0.0057          $   & $0.3184\pm 0.0056          $\\
$\sigma_8                  $                     & $0.799^{+0.017}_{-0.023}   $ & $0.8144\pm 0.0087          $   & $0.8066\pm 0.0083          $\\
${\rm{Age}}/\mathrm{Gyr}   $                     & $13.762\pm 0.021           $ & $13.755\pm 0.020           $   & $13.759\pm 0.020           $\\
$r_\mathrm{drag}           $                     & $147.40\pm 0.22            $ & $147.43\pm 0.22            $   & $147.46\pm 0.21            $\\

        \hline
    \end{tabular}
\caption{Marginalized constraints at 68\% CL for the sampled and derived cosmological parameters within the Parabolic parameterization using three dataset combinations: CMB + DESI, CMB + DESI + Pantheon+, and CMB + DESI + DES-SN5YR.
}
\label{tab:Parabolic}
\end{table}

\begin{figure}[tbp]
    \centering
    \includegraphics[width=\textwidth]{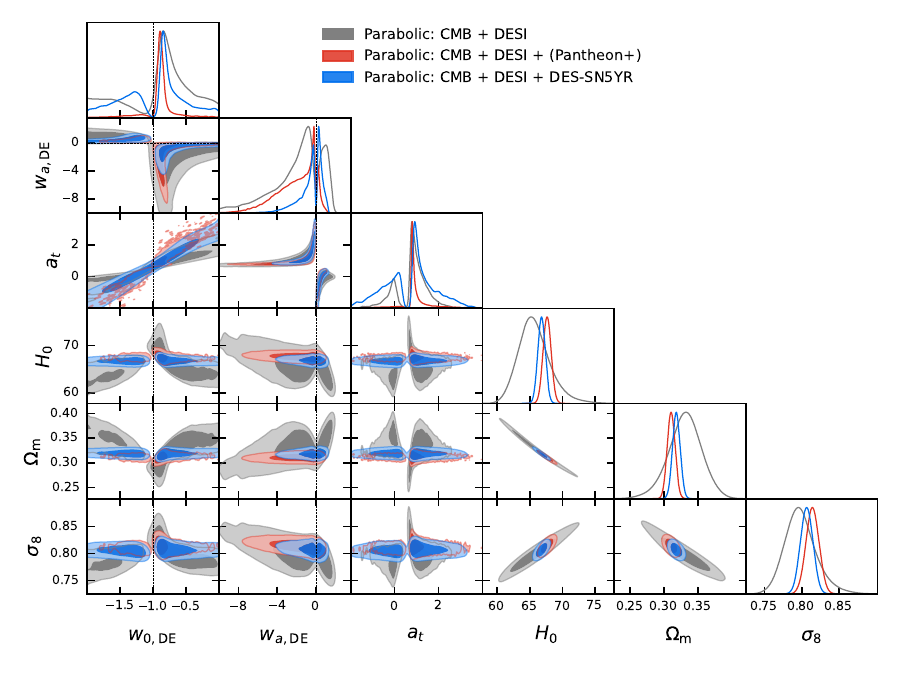}
    \caption{Triangle plot depicting 68\% and 95\% CL for select cosmological parameters alongside their 1D marginalized posterior distributions under the Parabolic parameterization. Contours are shown for three dataset combinations:1)CMB + DESI (gray), 2) CMB + DESI + Pantheon+ (red), and 3) CMB + DESI + DES-SN5YR (blue). The black dashed lines indicate the baseline $\Lambda\text{CDM}$ values, $(w_{0}, w_{a}) = (-1, 0)$.}
    \label{fig:parabolic}
\end{figure}

% ---------------------------
In Table ~\ref{tab:Parabolic}, the marginalized posterior constraints for main cosmological parameters and the parabolic model parameters are listed  at 68\% CL for the three dataset combinations. The corresponding 1D and 2D posterior distributions are shown in the triangular plot provided in Figure~\ref{fig:parabolic} where the intersection of the black dashed lines in ($w_{0,DE}$ and $w_{a,DE}$) plot corresponds to the baseline $\Lambda\text{CDM}$ values, $(w_{0}, w_{a}) = (-1, 0)$. Similar to CPL and Wang results, a combination of DESI BAO measurements with CMB data alone provides poor constraints on the Equation of State (EoS) parameters  (grey contours in Figure~\ref{fig:parabolic}):
% ---------------------------
\begin{subequations}\label{eq:eos_parabola_1}
\begin{align}
\label{eq:eos_parabola_1:w0}
w_{0} &  < -0.666    
\\
\label{eq:eos_parabola_1:wa}
w_{a} & = -1.9^{+3.4}_{-1.1} 
\\
\label{eq:eos_parabola_1:wat}
a_{t}  & =  0.78^{+0.65}_{-0.93}  
\end{align}
\end{subequations}
% ---------------------------
For the CMB + DESI + Pantheon+ combination, the marginalized posterior constraints for the EoS parameters are:
% ---------------------------
\begin{subequations}\label{eq:eos_parabola_2}
\begin{align}
\label{eq:eos_parabola_2:w0}
w_{0} & = -0.889^{+0.075}_{-0.068} 
\\
\label{eq:eos_parabola_2:wa}
w_{a} & =-2.1^{+2.5}_{-1.1} 
\\
\label{eq:eos_parabola_2:wat}
a_{t}  & =  0.93\pm 0.70 
\end{align}
\end{subequations}
% ---------------------------
Observing the error bars, only $w_{0}$ is relatively well-constrained. The turning point, $a_t$ for the EoS, however is occurred at $a \sim 0.9$ with very large uncertainty of 75\% at 68\% CL.

For the CMB + DESI + DES-SN5YR combination, the marginalized posterior constraints for the EoS parameters are:
% ---------------------------
\begin{subequations}\label{eq:eos_parabola_3}
\begin{align}
\label{eq:eos_parabola_3:w0}
w_{0} & = -0.99^{+0.43}_{-0.52} 
\\
\label{eq:eos_parabola_3:wa}
w_{a} & = -0.52^{+1.4}_{-0.54}
\\
\label{eq:eos_parabola_3:wat}
a_{t}  & =  0.7^{+1.1}_{-1.2}    
\end{align}
\end{subequations}
% ---------------------------

The EoS parameters ($w_0, w_a, a_t$) exhibit significant uncertainties at the 68\% CL. Despite the use of modern datasets, these constraints have not seen substantial improvement since earlier studies~\cite{Hu2014}. Nevertheless, we find a preferred turning point at a scale factor of $a_t \sim 0.7$ which closely aligns with the value reported in Ref.~\cite{Hu2014}.

Regarding the parabolic parameterization, previous constraints—utilizing a combination of SNLS3, Planck/WMAP9, and HST, alongside BAO data from 6dFGS, BOSS DR11, and WiggleZ—showed negligible evidence against the $\Lambda$CDM model~\cite{Hu2014}. In contrast, our current analysis, particularly the CMB + DESI + DES-SN5YR combination, reveals a substantial preference for a dynamical dark energy component at the $4\sigma$ significance level.

%===========================================

%===========================================

\begin{figure}[tbp]
    \centering
    \includegraphics[width=0.7\textwidth]{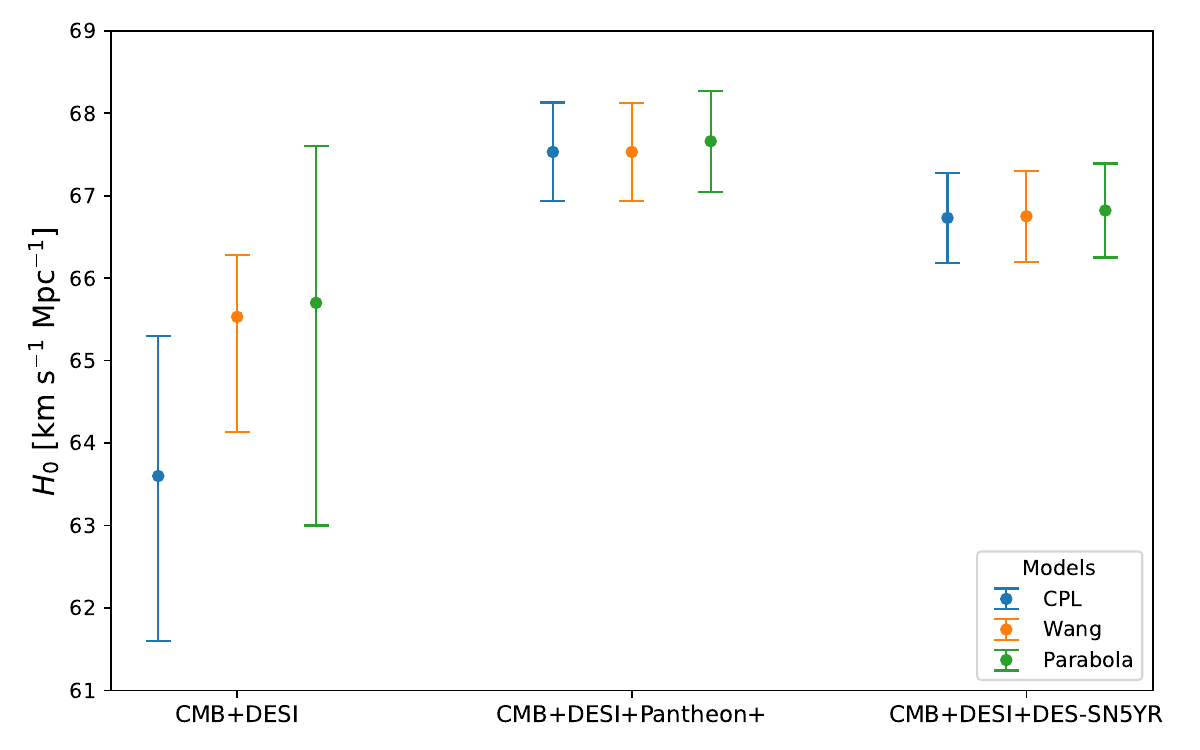}
    % \qquad
    \caption{ Marginalized posterior for the Hubble constant, $H_0$, at 68\% CL. The results are grouped by dataset combinations: 1) CMB + DESI, 2) CMB + DESI + Pantheon+, and 3) CMB + DESI + DES-SN5YR (ordered from left to right). Within each dataset cluster, the three considered dark energy models are plotted in close proximity to facilitate direct comparison of their impact on the $H_0$ estimate.}
    \label{fig:h0}
\end{figure}

%===========================================
\begin{table}[tbp]
    \centering
%    \begin{tabular}{|c|c|c|c|}
    \begin{tabular}{|c|c|c|c|c|c|}
%        \hline
        \hline
        \rule{0pt}{3ex}
        \bf Params.
        &$\Delta\chi^2              $
        &Significance
        &FoM 
        &$\Delta \text{AIC}$
        &$\Delta \text{BIC}$
        \\[0.5ex]
        \hline

%\hline
%        \rule{0pt}{3ex}
        % Using \multicolumn{1}{c}{} to remove the | for these specific cells
%        \multicolumn{1}{|c}{\rule{0pt}{3ex}}  
        \multicolumn{1}{|c}{}  
        & \multicolumn{3}{c}{\bf CMB+DESI} 
%        & \multicolumn{1}{c}{} 
        & \multicolumn{1}{c}{} 
        & \multicolumn{1}{c|}{} 
        \\ 
        \hline
        \rule{0pt}{3ex}

%-----------------------------------------------------------------------------------------------
CPL            & -10.5        &  2.8$\sigma$   &  40.50      & +6.5    & +7.92   \\
Wang           & -10.9        &  2.9$\sigma$   &  244.37     & +6.9    & +7.52   \\
Parabola       & -11.3        &  2.6$\sigma$   & 3.10        & +5.3    & +16.33       \\[0.5ex]

%-----------------------------------------------------------------------------------------------
        \hline
        \multicolumn{1}{|c}{} &        
        \multicolumn{3}{c}{\bf CMB+DESI+(Pantheon+)} 
%        & \multicolumn{1}{c}{} 
        & \multicolumn{1}{c}{} 
        & \multicolumn{1}{c|}{} 
        \\ 
        \hline

%-----------------------------------------------------------------------------------------------
CPL            & -8.6         &  2.5$\sigma$   &  193.34    & +4.6    & +10.13  \\
Wang           & -9.5         &  2.6$\sigma$   &  594.65    & +5.5    & +9.23   \\
Parabola       & -12.5        &  2.8$\sigma$   &  9.96      & +6.5    & +15.60  \\[0.5ex]

%-----------------------------------------------------------------------------------------------
         \hline
        \multicolumn{1}{|c}{} &        
        \multicolumn{3}{c}{\bf CMB+DESI+(DES-SN5YR)} 
%        & \multicolumn{1}{c}{} 
        & \multicolumn{1}{c}{} 
        & \multicolumn{1}{c|}{} 
        \\ 
        \hline

%-----------------------------------------------------------------------------------------------
CPL            &  -20.5     &  4.1$\sigma$   &  187.98    & +16.5    & -1.74  \\
Wang           &  -19.4     &  4.0$\sigma$   &  560.08    & +15.4    & -0.64   \\
Parabola       &  -21.7     &  4.0$\sigma$   &  7.10      & +15.7    & +6.43   \\[0.5ex]

%-----------------------------------------------------------------------------------------------
        \hline
    \end{tabular}
\caption{Statistical comparison of the dark energy parameterizations we considered across three dataset combinations: 1) CMB+DESI, 2) CMB+DESI+Pantheon+, and 3) CMB+DESI+DES-SN5YR. We report the goodness-of-fit relative to the $\Lambda$CDM model ($\Delta \chi^{2} \equiv \chi^{2}_{\text{model,i}} - \chi^{2}_{\Lambda\text{CDM}}$), the frequentist significance of the deviation from the $\Lambda$CDM model, the relative Figure of Merit (FoM), and the relative information criteria ($\Delta \text{AIC} = \text{AIC}_{\Lambda\text{CDM}} - \text{AIC}_{\text{model,i}}$ and $\Delta \text{BIC} = \text{BIC}_{\text{model,i}} - \text{BIC}_{\Lambda\text{CDM}}$). By definition, smaller values of $\chi^{2}$, $\text{AIC}$, and $\text{BIC}$ indicate a better-performing model.}

\label{tab:comparision}
\end{table}

%===========================================
In Figure \ref{fig:h0}, we demonstrate the $H_{0}$ posterior variation with respect to three different dataset combinations for the considered dark energy parameterizations. One can observe that the variation in the $H_{0}$ estimates is driven primarily by the dataset combination; specifically, the inclusion of supernova compilations exerts a distinct 'pull' toward higher $H_{0}$ values compared to the CMB+DESI baseline.

This shift in $H_{0}$ is closely reflected in the varying degrees of statistical evidence for dynamical dark energy across the different samples. As evidenced in Table~\ref{tab:comparision}, the CMB+DESI combination alone shows limited evidence for dynamical dark energy, with all models falling below the $3\sigma$ significance threshold. Interestingly, the inclusion of the Pantheon+ dataset results in a slightly weaker preference for the specific dynamical evolution described by these models. Conversely, a robust $4\sigma$ preference is achieved across all parameterizations when the DES-SN5YR sample is utilized in conjunction with the CMB+DESI baseline.

While looking at the Figure of Merit (FoM) column in Table \ref{tab:comparision}, the two-parameter $(w_{0}, w_{a})$ models---specifically CPL and Wang---the Figure of Merit (FoM) increases substantially upon the inclusion of the supernova datasets (for both Pantheon+ and DES-SN5YR).  Across all dataset combinations, the Wang model consistently achieves the highest FoM, despite maintaining a moderate $\Delta \chi^{2}$ improvement relative to the other parameterizations. 

Regarding the three-parameter Parabolic model, the Figure of Merit (FoM) is significantly lower than those of the two-parameter $(w_0, w_a)$ models, as anticipated. This reduction in the FoM value is a direct consequence of the expanded parameter space, i.e., an addition of $a_t$ to $(w_{0}, w_{a})$ increases the $N$-dimensional volume enclosed by the 68\% CL region. Despite this reduction in constraining power, the $\Delta \chi^{2}$ values for the Parabolic model remain comparable to the CPL and Wang parameterizations across all considered dataset combinations.

When we observe at AIC selection criteria, as shown in the column~5 of Table~\ref{tab:comparision}, $\Delta\text{AIC}$ values for all parameterizations we employed, i.e., $\Delta\text{AIC} > 4$ reveal weak (little) evidence in favor of the $\Lambda\text{CDM}$ model for the cases of the CMB + DESI and the CMB + DESI + Pantheon+ combinations. The dynamical models achieve the trend of $ \text{AIC}$ values order as: $ \text{Wang} < \text{CPL} < \text{Parabola} < \Lambda\text{CDM}$ and $\text{Parabola} < \text{Wang} < \text{CPL} <  \Lambda\text{CDM}$ with the CMB + DESI  and the CMB + DESI + Pantheon+ combinations respectively. Most notably, the data yields no evidence in favor of $\Lambda\text{CDM}$ over its dynamic counterparts when using the joint CMB + DESI + DES-SN5YR dataset since $\Delta\text{AIC} > 7$,  so as the alternative models  are very strongly favored with the following order: $\text{CPL} < \text{Parabola} < \text{Wang} < \Lambda\text{CDM}$.

However, when shifting the model comparison to the BIC framework, the conclusions alter substantially. Given our extensive data volume ($N \sim 10^4$), the BIC imposes a significantly heavier penalty on additional parameters than the AIC due to its logarithmic $\ln(N)$ scaling factor. Consequently, the $\Delta\text{BIC}$ values as shown in the last column of ~Table~\ref{tab:comparision} indicate "strong (decisive) evidence against" the dynamical dark energy parameterizations compared to the  $\Lambda\text{CDM}$ model for both the CMB + DESI and CMB + DESI + Pantheon+ combinations with the same order of $ \text{BIC}:  \Lambda\text{CDM} < \text{Wang} < \text{CPL} < \text{Parabola} $. Interestingly, on substituting the supernova DES-SN5YR dataset within the joint combination, the ordering of BIC shifts as $\text{BIC}:  \text{CPL} < \text{Wang} < \Lambda\text{CDM}<  \text{Parabola} $ and $\Delta\text{BIC}$ yields negligible evidence against the CPL and Wang models as their BIC values are lower when compared with the $\Lambda\text{CDM}$ value. The fluctuation in the relative ranking of the CPL and Wang models under AIC and BIC criteria across different Type Ia supernova (SNe~Ia) compilations can potentially be attributed to variations in their respective redshift coverages and data point distributions. As expected, due to the extra degree of freedom in the Parabolic model, its $\Delta\text{BIC}$ demonstrates "strong evidence against" it even within the DES-SN5YR-inclusive combination. This indicates that the added complexity of a third parameter fails to yield a sufficient improvement in $\chi^2_{\text{min}}$ to offset the severe parameter penalty, demonstrating that current data do not yet justify the additional degree of freedom.

%% file: Text/5_conclusions.tex
% !TEX root = ../main.tex

 \section{Summary and Conclusions}
\label{sec:conclusion}

The DESI collaboration’s recent BAO measurements, analyzed under the CPL parameterization, have sparked considerable interest due to their reports of $3.8\sigma$ to $4.2\sigma$ preference for the dynamical dark energy model~\cite{tr6y-kpc6}. In this paper, we have reproduced the similar results upon considering the CPL parameterization using the DESI BAO measurements, the cosmic microwave background and Type Ia supernovae data and the result again underscores the potential departure from the $\Lambda$CDM framework. Furthermore, in this study, we extend the CPL analysis by incorporating two alternative parameterizations of the dark energy equation of state $w(z)$: the two-parameter Wang model~\cite{wang}and a three-parameter parabolic model~\cite{Hu2014}. Utilizing the MCMC sampler \texttt{Cobaya} \cite{Torrado:2020dgo} integrated with the \texttt{CAMB} Boltzmann solver~\cite{Lewis:1999bs,Howlett:2012mh}, we placed constraints on both key cosmological parameters and the model parameters across three joint dataset combinations: 1) CMB + DESI, 2) CMB + DESI + Pantheon+, and 3) CMB + DESI + DES-SN5YR. Our primary findings, evaluated across the various model metrics, are summarized as follows:

\begin{itemize}

\item The CMB + DESI dataset combination provides least constraints on the parameters space as expected for all three parameterizations when compared with the datasets including the type Ia supernovae data: Pantheon+ and DES-SN5YR. Further we notice that in case of the parabolic model the constraints on EoS parameters exhibits significant uncertainties even with the use of recent datasets.

\item Across all three models, we find that the CMB + DESI and CMB + DESI + Pantheon+ combinations yield less than $3\sigma$ significance, providing the preference for the dynamical dark energy model upto $3 \sigma$ over the standard $ \Lambda$CDM model while the inclusion of DES-SN5YR increases this to $\sim4\sigma$. Therefore, we conclude that the inclusion of the DES-SN5YR dataset into the CMB + DESI combination strengthens the preference for dynamical dark energy models.

\item The Wang model consistently shows the smallest posterior volume (tight bound) for the EoS parameters for all the dataset combination, resulting in the highest relative Figure of Merit (FoM) with the order of: $\text{Wang} > \text{CPL} > \text{Parabolic}$.

\item Interestingly, the relative differences in the Akaike Information Criterion (AIC) indicate weak evidence in favor of the $\Lambda\text{CDM}$ model over the dynamical models we considered when using the CMB + DESI and CMB + DESI + Pantheon+ combinations and this evidence towards the $\Lambda\text{CDM}$ model vanishes when the joint CMB + DESI + DES-SN5YR dataset is utilized, thereby very strongly favoring the dynamical dark energy models.

\item While the aforementioned metrics hint at a preference for dynamical dark energy, the Bayesian Information Criterion (BIC) provides a more cautious perspective. Due to its dependence on the number of free parameters and the total number of data points used, BIC imposes a heavy penalty on these extra degrees of freedom, offering mild support for an evolving dark energy component---specifically the CPL and Wang models---over the standard $\Lambda\text{CDM}$ model  on utilizing the CMB + DESI + DES-SN5YR dataset. Consequently, $\Lambda\text{CDM}$ model remains the preferred, standard model according to BIC benchmarks.

\end{itemize}

Our finding of the $\text{N}\sigma$ and AIC values, appear to favor the three-parameter Parabolic model over $\Lambda\text{CDM}$ but there is no significant improvement in its minimum $\chi^2$, despite the addition of a third degree of freedom ($a_t$). Instead, the expanded parameter space inflates the $N$-dimensional volume of the $68\%$ confidence region, severely degrading the dark energy figure of merit. In addition, its $\Delta\text{BIC}$ values indicate "strong evidence against" the Parabolic model across the considered datasets, demonstrating that the current cosmological data do not  justify an introduction of higher-order parameterization of dark energy EoS. However, upcoming high and intermediate redshift BAO measurements will extend the current dataset, providing a more rigorous viability test for these cosmological models
    
Finally, we conclude from our study that two-parameter descriptions of $w(z)$, such as the CPL and Wang models, remain the optimal choice to describe the dynamical nature of 
dark energy with the current available data .

%% file: Text/6_appendix_w_f.tex
% !TEX root = ../main.tex

\section{Appendix: Posterior Distributions for $w(z)$ and $F(z)$}
\label{sec:w_f}

\begin{figure}[tbp]
    \centering
    \includegraphics[width=0.32 \textwidth]{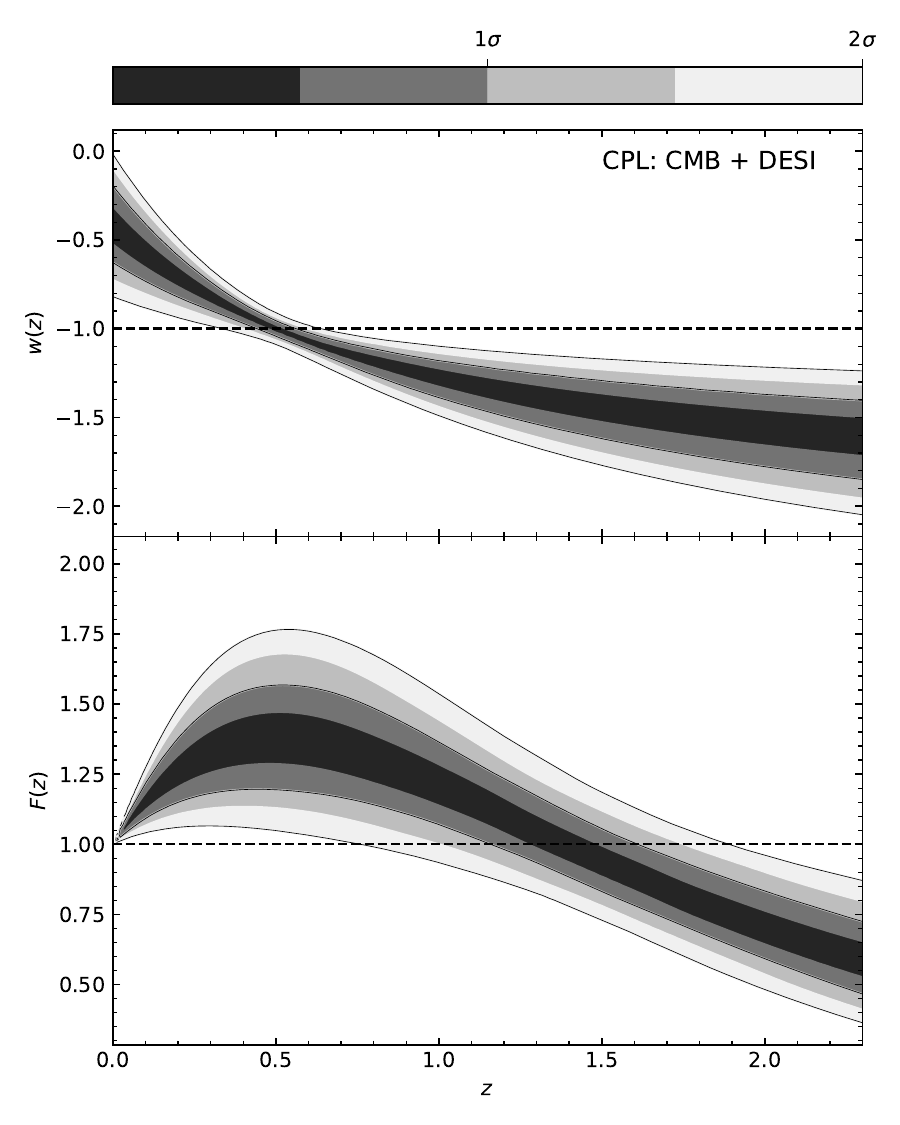}
    \includegraphics[width=0.32 \textwidth]{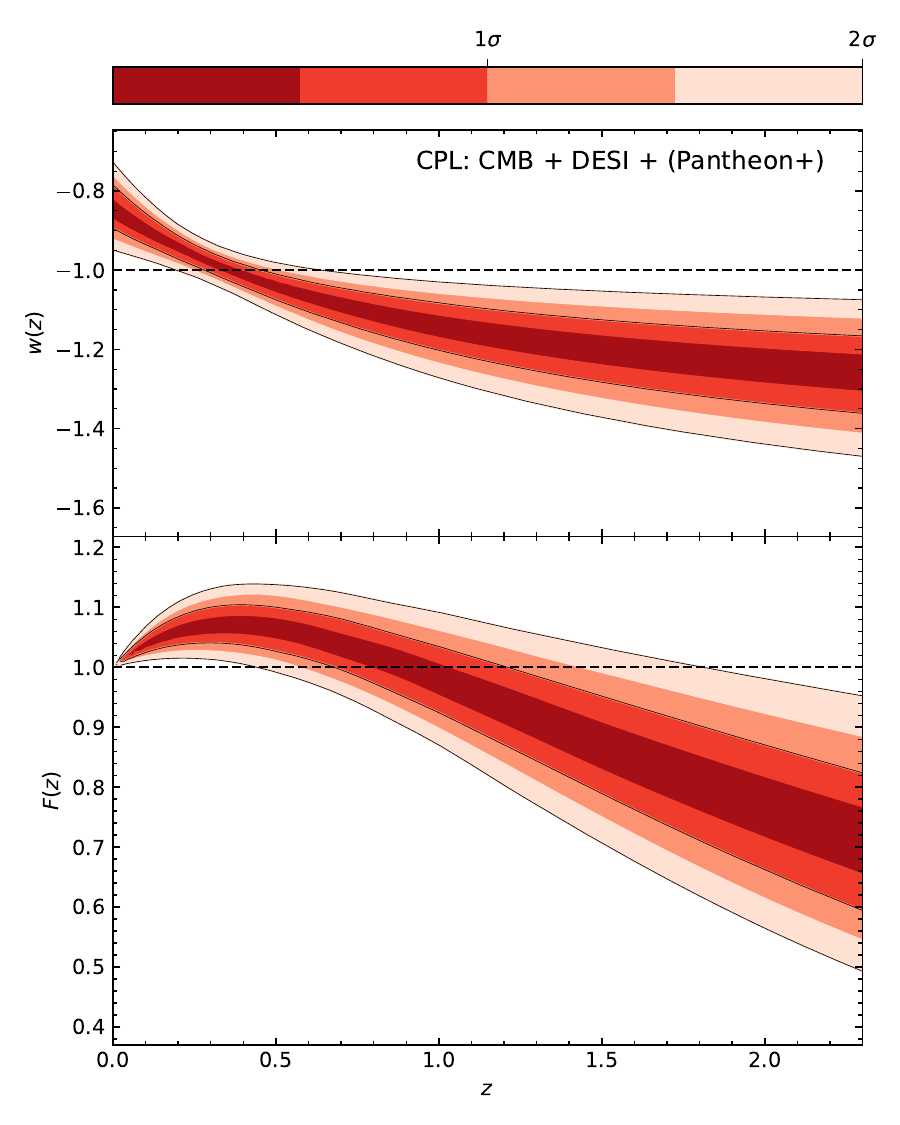}
    \includegraphics[width=0.32 \textwidth]{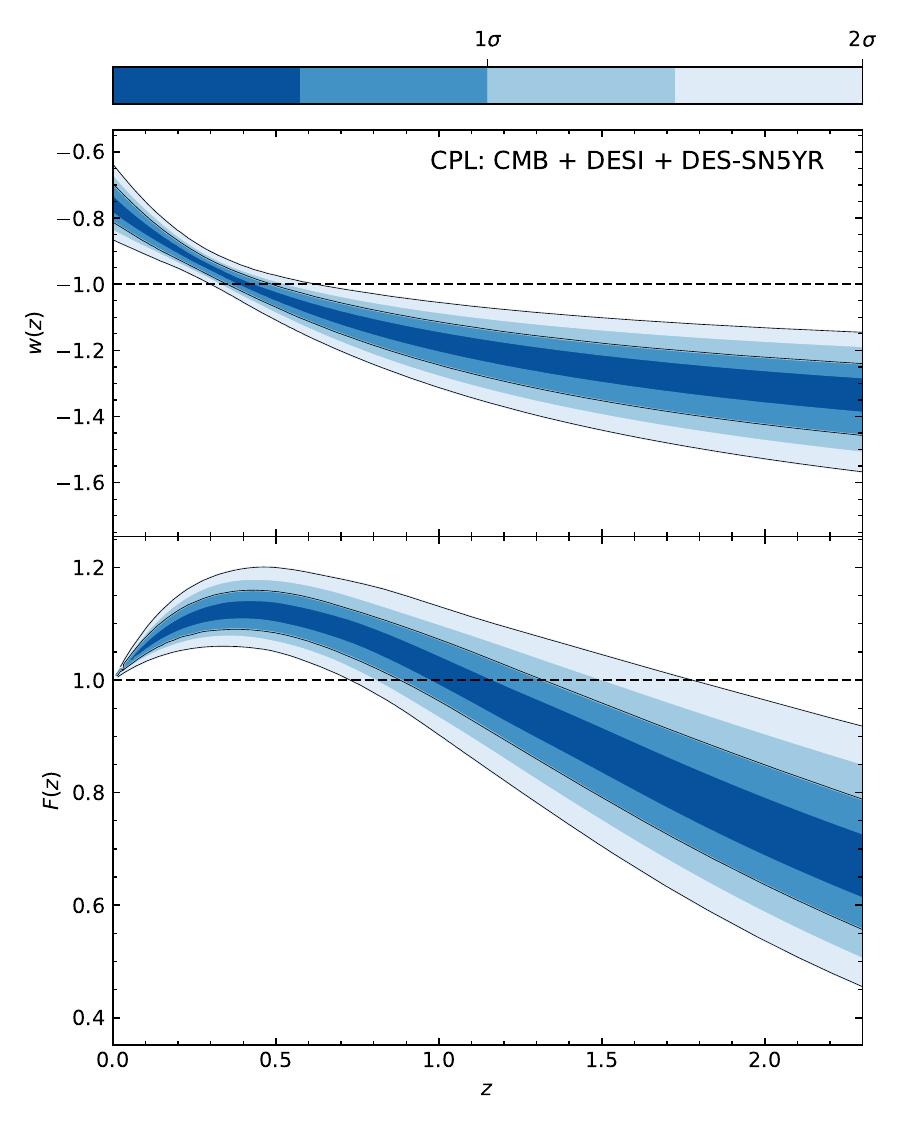}
    
    \includegraphics[width=0.32 \textwidth]{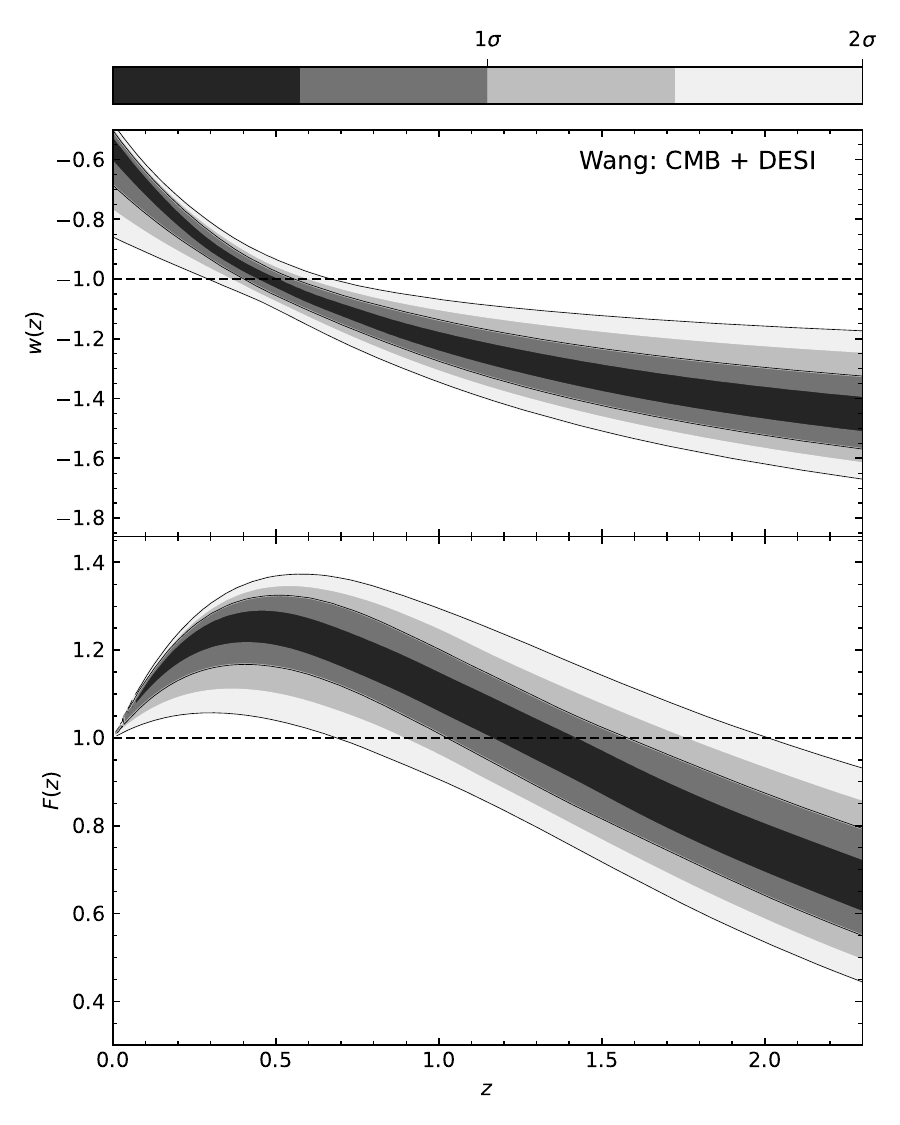}
    \includegraphics[width=0.32 \textwidth]{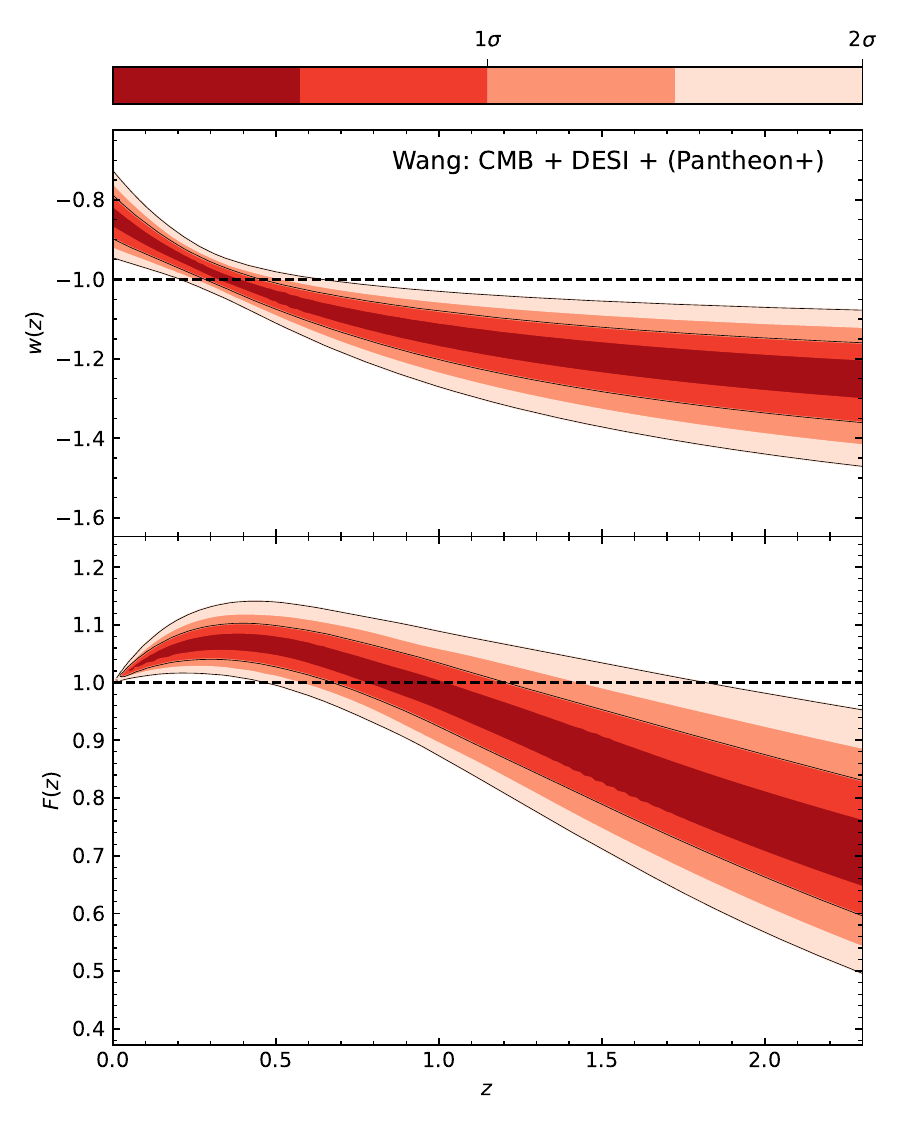}
    \includegraphics[width=0.32 \textwidth]{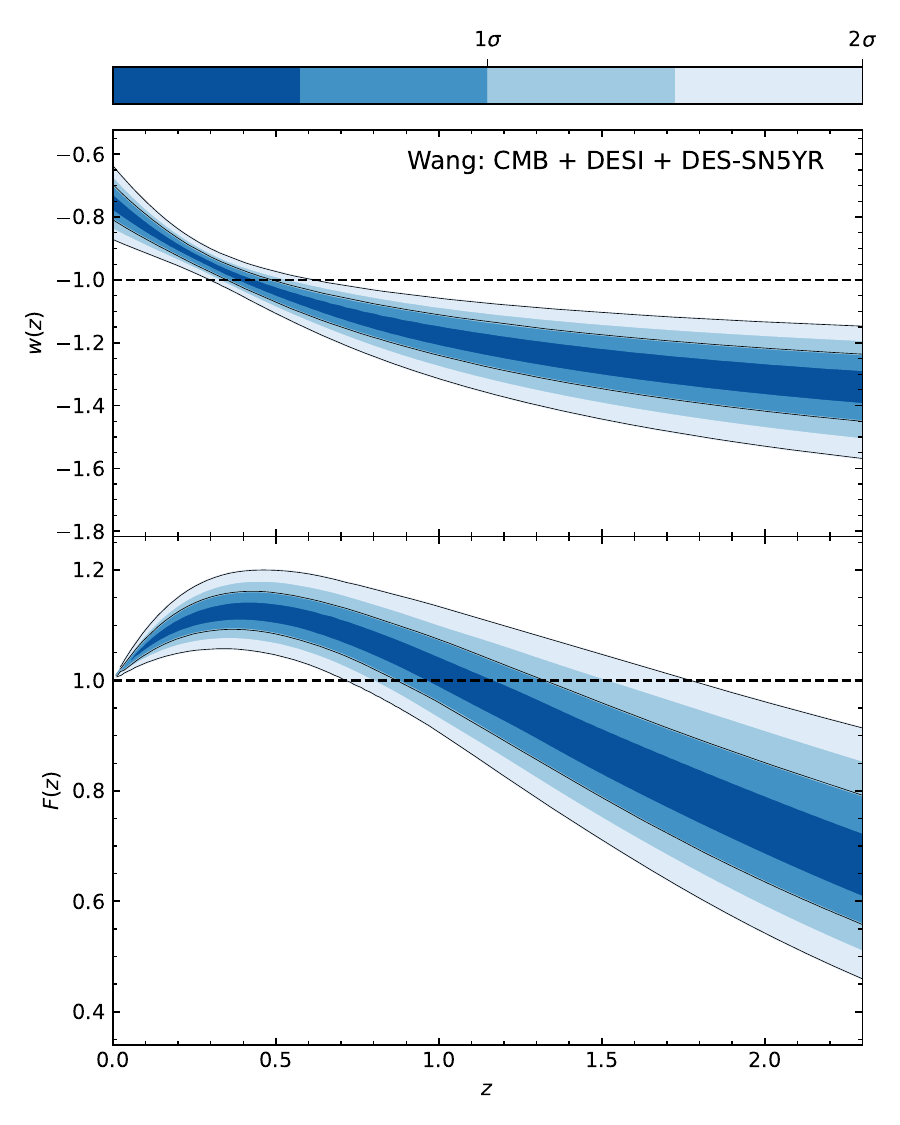}        
    \includegraphics[width=0.32 \textwidth]{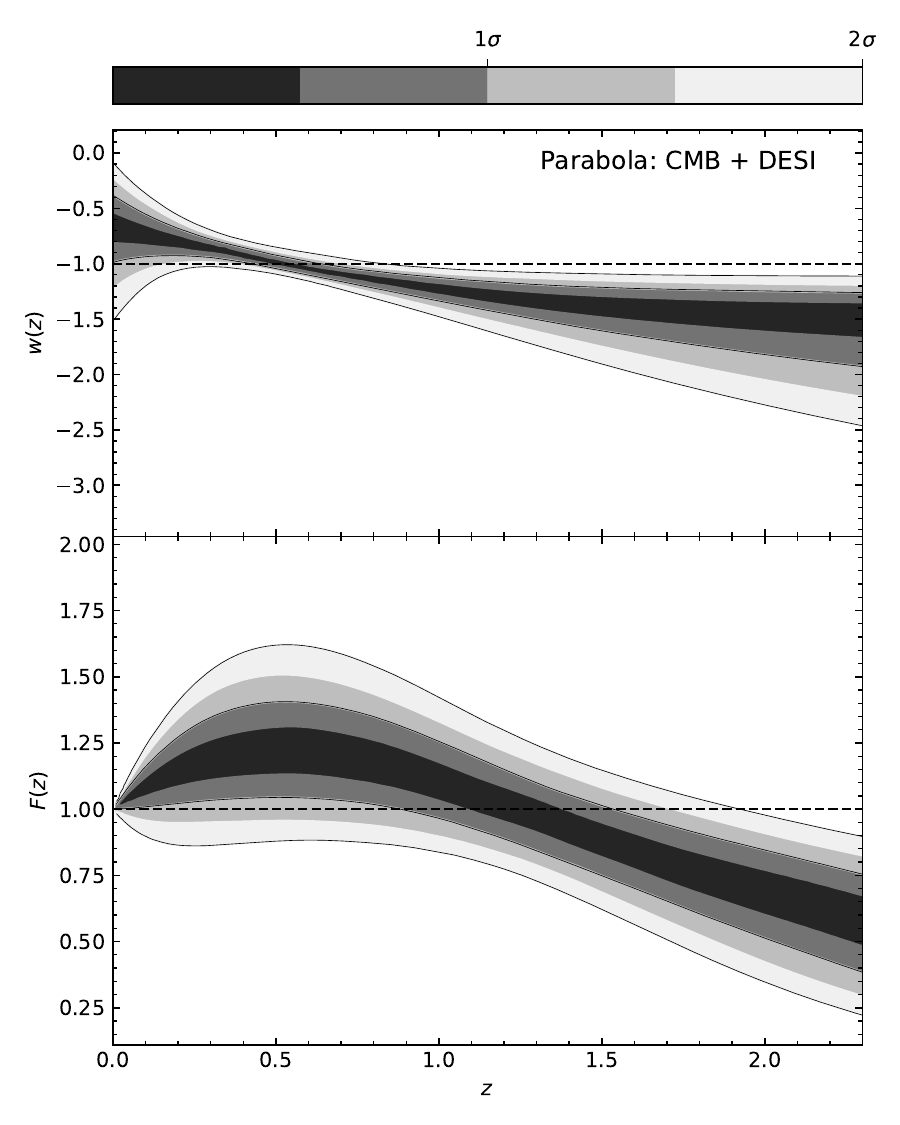}    \includegraphics[width=0.32 \textwidth]{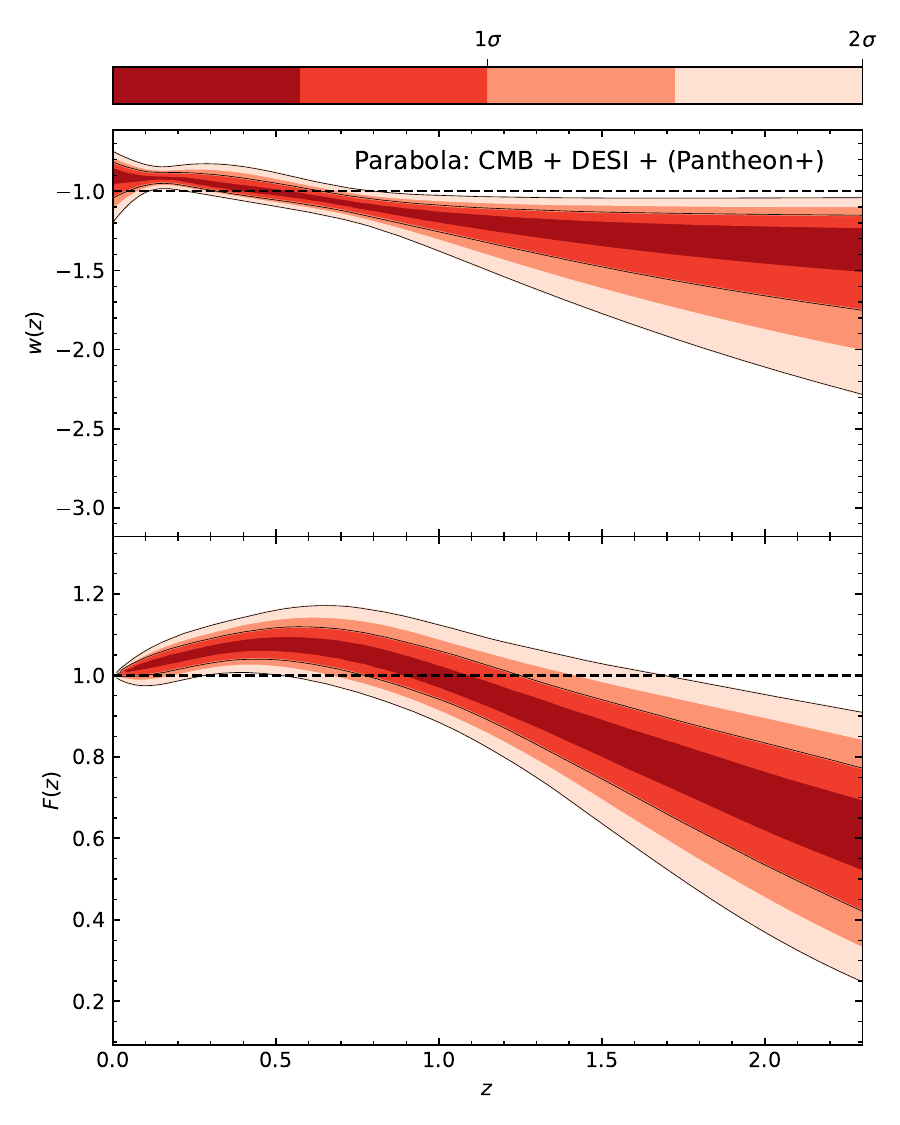}
    \includegraphics[width=0.32 \textwidth]{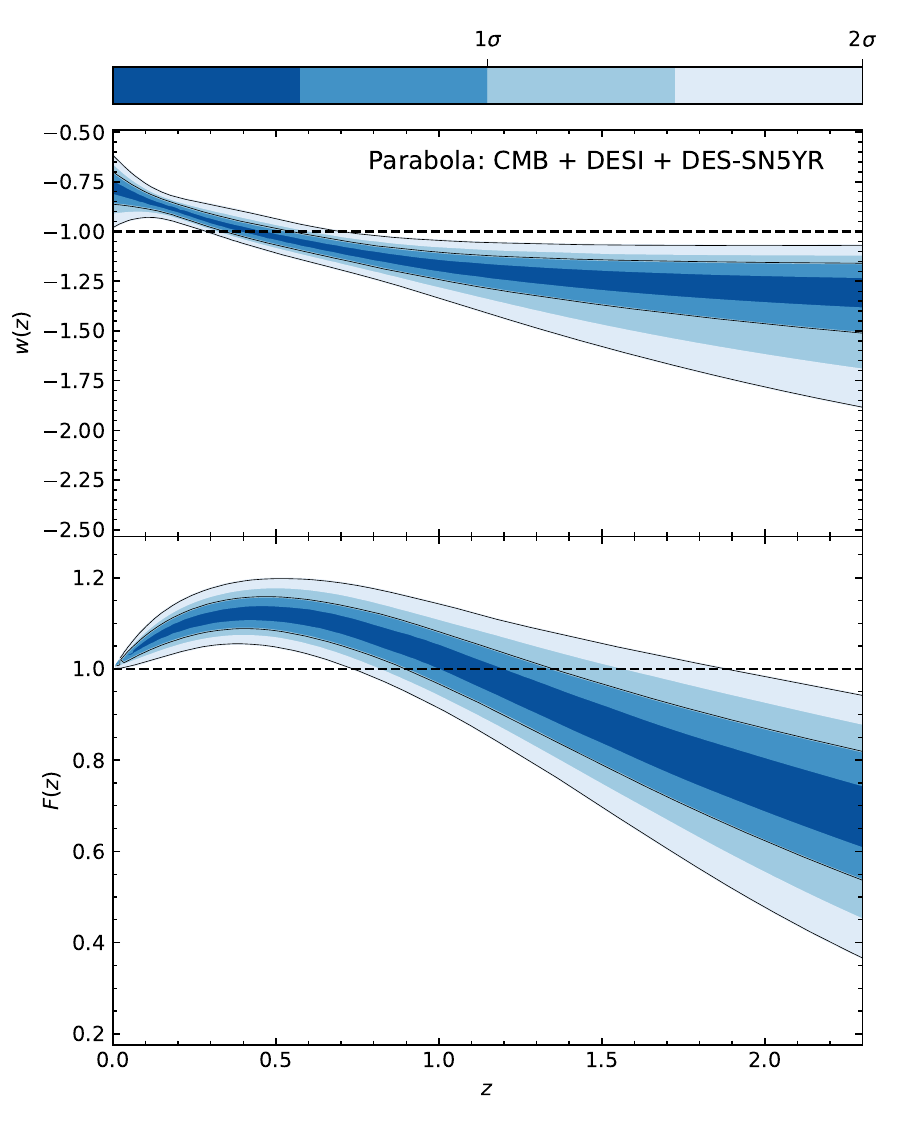}        % \qquad

\caption{Reconstructed posterior distributions for the Equation of State (EoS) and $F(z)$ (defined in Eq.~\eqref{eq:f}) at $1\sigma$ and $2\sigma$ confidence levels (CL). Results are shown for the CPL, Modified CPL (Wang), and Parabolic parameterizations (from top to bottom) using 1) CMB + DESI (gray),  2) CMB + DESI + (Pantheon+)(red), and  3) CMB + DESI + DES-SN5YR(blue) dataset combinations. 
The black dashed lines in each plot correspond to the $\Lambda$CDM model.
}

    \label{fig:wos_F_against_z}
\end{figure}

In this section, we present the constraints on the evolution of the dark energy equation of state $w(z)$ and the fractional dark energy density, $F(z)$ defined in eq.~\eqref{eq:f}. To generate the confidence contours shown in Fig.~\ref{fig:wos_F_against_z}, we employ the publicly available \texttt{fgivenx} package~\cite{fgivenx}\footnote{\url{https://github.com/handley-lab/fgivenx}}. These contours illustrate the redshift dependence of the $w(z)$, and the relative dark energy density, $F(z)$ at $1\sigma$ and $2\sigma$ confidence levels (CL) for the CPL, Wang, and Parabolic parameterizations using 3 dataset combinations 1)CMB + DESI (gray), 2)CMB + DESI + (Pantheon+)(red), and 3) CMB + DESI + DES-SN5YR (blue). The black dashed lines in Fig.~\ref{fig:wos_F_against_z}    correspond to the $\Lambda$CDM model.

Across all considered models and dataset combinations, we consistently observe that a phantom crossing ($w(z) = -1$) occurs at a redshift of $z \sim 0.5$, very close to the point where the relative dark energy density begins to decline. Coincidentally, we notice that around this redshift regime the equation of state, $w(z)$, is most tightly constrained by the data we used. Interestingly, even though the CMB + DESI combination alone does provide a loose bound on cosmological and model parameters, it exhibits an evolutionary trend of $w(z)$ and $F(z)$ remarkably similar to the other two dataset combinations that incorporate Type Ia supernovae.

%% file: references.bib
@article{Riess_1998,
doi = {10.1086/300499},
url = {https://doi.org/10.1086/300499},
year = {1998},
month = {sep},
publisher = {},
volume = {116},
number = {3},
pages = {1009},
author = {Riess, Adam G. and Filippenko, Alexei V. and Challis, Peter and Clocchiatti, Alejandro and Diercks, Alan and Garnavich, Peter M. and Gilliland, Ron L. and Hogan, Craig J. and Jha, Saurabh and Kirshner, Robert P. and Leibundgut, B. and Phillips, M. M. and Reiss, David and Schmidt, Brian P. and Schommer, Robert A. and Smith, R. Chris and Spyromilio, J. and Stubbs, Christopher and Suntzeff, Nicholas B. and Tonry, John},
title = {Observational Evidence from Supernovae for an Accelerating Universe and a Cosmological Constant},
journal = {The Astronomical Journal}
}

@article{Perlmutter_1999,
doi = {10.1086/307221},
url = {https://doi.org/10.1086/307221},
year = {1999},
month = {jun},
publisher = {},
volume = {517},
number = {2},
pages = {565},
author = {Perlmutter, S. and Aldering, G. and Goldhaber, G. and Knop, R. A. and Nugent, P. and Castro, P. G. and Deustua, S. and Fabbro, S. and Goobar, A. and Groom, D. E. and Hook, I. M. and Kim, A. G. and Kim, M. Y. and Lee, J. C. and Nunes, N. J. and Pain, R. and Pennypacker, C. R. and Quimby, R. and Lidman, C. and Ellis, R. S. and Irwin, M. and McMahon, R. G. and Ruiz-Lapuente, P. and Walton, N. and Schaefer, B. and Boyle, B. J. and Filippenko, A. V. and Matheson, T. and Fruchter, A. S. and Panagia, N. and Newberg, H. J. M. and Couch, W. J. and Project, The Supernova Cosmology},
title = {{Measurements of $\Omega$ and $\Lambda$ from 42 High Redshift Supernovae}},
journal = {The Astrophysical Journal}
}

@article{Copeland,
author = {Copeland, Edmund J. and Sami, M. and Tsujikawa, Shinji},
title = {DYNAMICS OF DARK ENERGY},
journal = {International Journal of Modern Physics D},
volume = {15},
number = {11},
pages = {1753-1935},
year = {2006},
doi = {10.1142/S021827180600942X},
URL = {https://doi.org/10.1142/S021827180600942X},
xeprint = {https://doi.org/10.1142/S021827180600942X}
}

@article{BarbozaAlcaniz,
    author = "Barboza, Jr., E. M. and Alcaniz, J. S.",
    title = "{A parametric model for dark energy}",
    eprint = "0805.1713",
    archivePrefix = "arXiv",
    primaryClass = "astro-ph",
    doi = "10.1016/j.physletb.2008.08.012",
    journal = "Phys. Lett. B",
    volume = "666",
    pages = "415--419",
    year = "2008"
}

@article{JassalBaglaPadmanabhan,
    author = "Jassal, Harvinder Kaur and Bagla, J. S. and Padmanabhan, T.",
    title = "{Observational constraints on low redshift evolution of dark energy: How consistent are different observations?}",
    eprint = "astro-ph/0506748",
    archivePrefix = "arXiv",
    doi = "10.1103/PhysRevD.72.103503",
    journal = "Phys. Rev. D",
    volume = "72",
    pages = "103503",
    year = "2005"
}

@article{Dimakis,
    author = "Dimakis, N. and Karagiorgos, A. and Zampeli, Adamantia and Paliathanasis, Andronikos and Christodoulakis, T. and Terzis, Petros A.",
    title = "{General Analytic Solutions of Scalar Field Cosmology with Arbitrary Potential}",
    eprint = "1604.05168",
    archivePrefix = "arXiv",
    primaryClass = "gr-qc",
    doi = "10.1103/PhysRevD.93.123518",
    journal = "Phys. Rev. D",
    volume = "93",
    number = "12",
    pages = "123518",
    year = "2016"
}

@article{Pan,
    author = "Pan, Supriya and Yang, Weiqiang and Paliathanasis, Andronikos",
    title = "{Imprints of an extended Chevallier\textendash{}Polarski\textendash{}Linder parametrization on the large scale of our universe}",
    eprint = "1902.07108",
    archivePrefix = "arXiv",
    primaryClass = "astro-ph.CO",
    doi = "10.1140/epjc/s10052-020-7832-y",
    journal = "Eur. Phys. J. C",
    volume = "80",
    number = "3",
    pages = "274",
    year = "2020"
}

@article{tr6y-kpc6,
  title = {{DESI DR2 results. II. Measurements of baryon acoustic oscillations and cosmological constraints}},
  author = {Abdul Karim, M. and Aguilar, J. and Ahlen, S. and Alam, S. and Allen, L. and Prieto, C. Allende and Alves, O. and Anand, A. and Andrade, U. and Armengaud, E. and Aviles, A. and Bailey, S. and Baltay, C. and Bansal, P. and Bault, A. and Behera, J. and BenZvi, S. and Bianchi, D. and Blake, C. and Brieden, S. and Brodzeller, A. and Brooks, D. and Buckley-Geer, E. and Burtin, E. and Calderon, R. and Canning, R. and Rosell, A. Carnero and Carrilho, P. and Casas, L. and Castander, F. J. and Charles, M. and Chaussidon, E. and Chaves-Montero, J. and Chebat, D. and Chen, X. and Claybaugh, T. and Cole, S. and Cooper, A. P. and Cuceu, A. and Dawson, K. S. and de la Macorra, A. and de Mattia, A. and Deiosso, N. and Della Costa, J. and Demina, R. and Dey, A. and Dey, B. and Ding, Z. and Doel, P. and Edelstein, J. and Eisenstein, D. J. and Elbers, W. and Fagrelius, P. and Fanning, K. and Fern\'andez-Garc\'{\i}a, E. and Ferraro, S. and Font-Ribera, A. and Forero-Romero, J. E. and Frenk, C. S. and Garcia-Quintero, C. and Garrison, L. H. and Gazta\~naga, E. and Gil-Mar\'{\i}n, H. and Gontcho, S. Gontcho A. and Gonzalez, D. and Gonzalez-Morales, A. X. and Gordon, C. and Green, D. and Gutierrez, G. and Guy, J. and Hadzhiyska, B. and Hahn, C. and He, S. and Herbold, M. and Herrera-Alcantar, H. K. and Ho, M.-F. and Honscheid, K. and Howlett, C. and Huterer, D. and Ishak, M. and Juneau, S. and Kamble, N. V. and Kara\ifmmode \mbox{\c{c}}\else \c{c}\fi{}ayl��, N. G. and Kehoe, R. and Kent, S. and Kim, A. G. and Kirkby, D. and Kisner, T. and Koposov, S. E. and Kremin, A. and Krolewski, A. and Lahav, O. and Lamman, C. and Landriau, M. and Lang, D. and Lasker, J. and Le Goff, J. M. and Le Guillou, L. and Leauthaud, A. and Levi, M. E. and Li, Q. and Li, T. S. and Lodha, K. and Lokken, M. and Lozano-Rodr\'{\i}guez, F. and Magneville, C. and Manera, M. and Martini, P. and Matthewson, W. L. and Meisner, A. and Mena-Fern\'andez, J. and Menegas, A. and Mergulh\~ao, T. and Miquel, R. and Moustakas, J. and Mu\~noz-Guti\'errez, A. and Mu\~noz-Santos, D. and Myers, A. D. and Nadathur, S. and Naidoo, K. and Napolitano, L. and Newman, J. A. and Niz, G. and Noriega, H. E. and Paillas, E. and Palanque-Delabrouille, N. and Pan, J. and Peacock, J. A. and Ibanez, M. P. and Percival, W. J. and P\'erez-Fern\'andez, A. and P\'erez-R\`afols, I. and Pieri, M. M. and Poppett, C. and Prada, F. and Rabinowitz, D. and Raichoor, A. and Ram\'{\i}rez-P\'erez, C. and Rashkovetskyi, M. and Ravoux, C. and Rich, J. and Rocher, A. and Rockosi, C. and Rohlf, J. and Rom\'an-Herrera, J. O. and Ross, A. J. and Rossi, G. and Ruggeri, R. and Ruhlmann-Kleider, V. and Samushia, L. and Sanchez, E. and Sanders, N. and Schlegel, D. and Schubnell, M. and Seo, H. and Shafieloo, A. and Sharples, R. and Silber, J. and Sinigaglia, F. and Sprayberry, D. and Tan, T. and Tarl\'e, G. and Taylor, P. and Turner, W. and Ure\~na-L\'opez, L. A. and Vaisakh, R. and Valdes, F. and Valogiannis, G. and Vargas-Maga\~na, M. and Verde, L. and Walther, M. and Weaver, B. A. and Weinberg, D. H. and White, M. and Wolfson, M. and Y\`eche, C. and Yu, J. and Zaborowski, E. A. and Zarrouk, P. and Zhai, Z. and Zhang, H. and Zhao, C. and Zhao, G. B. and Zhou, R. and Zou, H.},
  collaboration = {DESI},
  journal = {Phys. Rev. D},
  volume = {112},
  issue = {8},
  pages = {083515},
  numpages = {40},
  year = {2025},
  month = {Oct},
  publisher = {American Physical Society},
  doi = {10.1103/tr6y-kpc6},
  url = {https://link.aps.org/doi/10.1103/tr6y-kpc6}
}

@article{Giare_2024,
doi = {10.1088/1475-7516/2024/10/035},
url = {https://dx.doi.org/10.1088/1475-7516/2024/10/035},
year = {2024},
month = {oct},
publisher = {IOP Publishing},
volume = {2024},
number = {10},
pages = {035},
author = {Giarè, William and Najafi, Mahdi and Pan, Supriya and Di Valentino, Eleonora and Firouzjaee, Javad T.},
title = {{Robust preference for Dynamical Dark Energy in DESI BAO and SN measurements}},
journal = {Journal of Cosmology and Astroparticle Physics}
}

@article{lodha2025extendeddarkenergyanalysis,
  title = {{Extended dark energy analysis using DESI DR2 BAO measurements}},
  author = {Lodha, K. and Calderon, R. and Matthewson, W. L. and Shafieloo, A. and Ishak, M. and Pan, J. and Garcia-Quintero, C. and Huterer, D. and Valogiannis, G. and Ure\~na-L\'opez, L. A. and Kamble, N. V. and Parkinson, D. and Kim, A. G. and Zhao, G. B. and Cervantes-Cota, J. L. and Rohlf, J. and Lozano-Rodr\'{\i}guez, F. and Rom\'an-Herrera, J. O. and Abdul-Karim, M. and Aguilar, J. and Ahlen, S. and Alves, O. and Andrade, U. and Armengaud, E. and Aviles, A. and Behera, J. and BenZvi, S. and Bianchi, D. and Brodzeller, A. and Brooks, D. and Burtin, E. and Canning, R. and Rosell, A. Carnero and Casas, L. and Castander, F. J. and Charles, M. and Chaussidon, E. and Chaves-Montero, J. and Chebat, D. and Claybaugh, T. and Cole, S. and Cuceu, A. and Dawson, K. S. and de la Macorra, A. and de Mattia, A. and Deiosso, N. and Demina, R. and Dey, Arjun and Dey, Biprateep and Ding, Z. and Doel, P. and Eisenstein, D. J. and Elbers, W. and Ferraro, S. and Font-Ribera, A. and Forero-Romero, J. E. and Garrison, Lehman H. and Gazta\~naga, E. and Gil-Mar\'{\i}n, H. and Gontcho, S. Gontcho A. and Gonzalez-Morales, A. X. and Gutierrez, G. and Guy, J. and Hahn, C. and Herbold, M. and Herrera-Alcantar, H. K. and Honscheid, K. and Howlett, C. and Juneau, S. and Kehoe, R. and Kirkby, D. and Kisner, T. and Kremin, A. and Lahav, O. and Lamman, C. and Landriau, M. and Le Guillou, L. and Leauthaud, A. and Levi, M. E. and Li, Q. and Magneville, C. and Manera, M. and Martini, P. and Meisner, A. and Mena-Fern\'andez, J. and Miquel, R. and Moustakas, J. and Santos, D. Mu\~noz and Mu\~noz-Guti\'errez, A. and Myers, A. D. and Nadathur, S. and Niz, G. and Noriega, H. E. and Paillas, E. and Palanque-Delabrouille, N. and Percival, W. J. and Pieri, Matthew M. and Poppett, C. and Prada, F. and P\'erez-Fern\'andez, A. and P\'erez-R\`afols, I. and Ram\'{\i}rez-P\'erez, C. and Rashkovetskyi, M. and Ravoux, C. and Ross, A. J. and Rossi, G. and Ruhlmann-Kleider, V. and Samushia, L. and Sanchez, E. and Schlegel, D. and Schubnell, M. and Seo, H. and Sinigaglia, F. and Sprayberry, D. and Tan, T. and Tarl\'e, G. and Taylor, P. and Turner, W. and Vargas-Maga\~na, M. and Walther, M. and Weaver, B. A. and Wolfson, M. and Y\`eche, C. and Zarrouk, P. and Zhou, R. and Zou, H.},
  collaboration = {DESI},
  journal = {Phys. Rev. D},
  volume = {112},
  issue = {8},
  pages = {083511},
  numpages = {27},
  year = {2025},
  month = {Oct},
  publisher = {American Physical Society},
  doi = {10.1103/w4c6-1r5j},
  url = {https://link.aps.org/doi/10.1103/w4c6-1r5j}
}

@article{Wolf_2025,
doi = {10.1088/1475-7516/2025/05/034},
url = {https://doi.org/10.1088/1475-7516/2025/05/034},
year = {2025},
month = {may},
publisher = {IOP Publishing},
volume = {2025},
number = {05},
pages = {034},
author = {Wolf, William J. and García-García, Carlos and Ferreira, Pedro G.},
title = {Robustness of dark energy phenomenology across different parameterizations},
journal = {Journal of Cosmology and Astroparticle Physics}
}

@article{ChevallierPolarski2001,
author = {Chevallier, Michel and Polarski, David},
title = {ACCELERATING UNIVERSES WITH SCALING DARK MATTER},
journal = {International Journal of Modern Physics D},
volume = {10},
number = {02},
pages = {213-223},
year = {2001},
doi = {10.1142/S0218271801000822},
URL = {https://doi.org/10.1142/S0218271801000822},
eprint = {gr-qc/0009008}
}

@article{Linder2003,
  title = {Exploring the Expansion History of the Universe},
  author = {Linder, Eric V.},
  journal = {Phys. Rev. Lett.},
  volume = {90},
  issue = {9},
  pages = {091301},
  numpages = {4},
  year = {2003},
  month = {Mar},
  publisher = {American Physical Society},
  doi = {10.1103/PhysRevLett.90.091301},
  url = {https://link.aps.org/doi/10.1103/PhysRevLett.90.091301},
  eprint = {astro-ph/0208512}
}

@article{fgivenx,
    doi = {10.21105/joss.00849},
    url = {http://dx.doi.org/10.21105/joss.00849},
    year  = {2018},
    month = {Aug},
    publisher = {The Open Journal},
    volume = {3},
    number = {28},
    author = {Will Handley},
    title = {fgivenx: Functional Posterior Plotter},
    journal = {The Journal of Open Source Software}
}

@ARTICLE{Akaike:1974,
  author={Akaike, H.},
  journal={IEEE Transactions on Automatic Control}, 
  title={A new look at the statistical model identification}, 
  year={1974},
  volume={19},
  number={6},
  pages={716-723},
  doi={10.1109/TAC.1974.1100705}}

@article{BIC:schwarz1978estimating,
  title={Estimating the dimension of a model},
  author={Schwarz, Gideon},
  journal={The annals of statistics},
  pages={461--464},
  year={1978},
  publisher={JSTOR}
}

@article{Liddle:2004,
    author = {Liddle, Andrew R.},
    title = {How many cosmological parameters},
    journal = {Monthly Notices of the Royal Astronomical Society},
    volume = {351},
    number = {3},
    pages = {L49-L53},
    year = {2004},
    month = {07},
    issn = {0035-8711},
    doi = {10.1111/j.1365-2966.2004.08033.x},
    url = {https://doi.org/10.1111/j.1365-2966.2004.08033.x},
    xeprint = {https://academic.oup.com/mnras/article-pdf/351/3/L49/3604253/351-3-L49.pdf},
}

@article{Liddle:2007,
    author = {Liddle, Andrew R.},
    title = {Information criteria for astrophysical model selection},
    journal = {Monthly Notices of the Royal Astronomical Society: Letters},
    volume = {377},
    number = {1},
    pages = {L74-L78},
    year = {2007},
    month = {05},
    issn = {1745-3925},
    doi = {10.1111/j.1745-3933.2007.00306.x},
    url = {https://doi.org/10.1111/j.1745-3933.2007.00306.x},
    xeprint = {https://academic.oup.com/mnrasl/article-pdf/377/1/L74/54686908/mnrasl\_377\_1\_l74.pdf},
}

@Article{Arevalo2017,
author={Arevalo, Fabiola
and Cid, Antonella
and Moya, Jorge},
title={{AIC and BIC for cosmological interacting scenarios}},
journal={The European Physical Journal C},
year={2017},
month={Aug},
day={21},
volume={77},
number={8},
pages={565},
issn={1434-6052},
doi={10.1140/epjc/s10052-017-5128-7},
url={https://doi.org/10.1140/epjc/s10052-017-5128-7}
}

@article{Forconietal2025,
  title = {Illustrating the consequences of a misuse of ${\ensuremath{\sigma}}_{8}$ in cosmology},
  author = {Forconi, Matteo and Favale, Arianna and G\'omez-Valent, Adri\`a},
  journal = {Phys. Rev. D},
  volume = {112},
  issue = {2},
  pages = {023517},
  numpages = {23},
  year = {2025},
  month = {Jul},
  publisher = {American Physical Society},
  doi = {10.1103/rpf5-ldks},
  url = {https://link.aps.org/doi/10.1103/rpf5-ldks}
}

@article{Aghanim:2019ame,
    author = "Aghanim, N. and others",
    collaboration = "Planck",
    title = "{Planck 2018 results. V. CMB power spectra and likelihoods}",
    eprint = "1907.12875",
    archivePrefix = "arXiv",
    primaryClass = "astro-ph.CO",
    doi = "10.1051/0004-6361/201936386",
    journal = "Astron. Astrophys.",
    volume = "641",
    pages = "A5",
    year = "2020"
}

@ARTICLE{rosenberg:2022,
       author = {{Rosenberg}, Erik and {Gratton}, Steven and {Efstathiou}, George},
        title = "{CMB power spectra and cosmological parameters from Planck PR4 with CamSpec}",
      journal = {\mnras},
         year = 2022,
        month = dec,
        volume = {517},
        number = {3},
        pages = {4620-4636},
          doi = {10.1093/mnras/stac2744},
archivePrefix = {arXiv},
       eprint = {2205.10869},
 primaryClass = {astro-ph.CO}
}

@article{Carron_2022,
doi = {10.1088/1475-7516/2022/09/039},
url = {https://doi.org/10.1088/1475-7516/2022/09/039},
year = {2022},
month = {sep},
publisher = {IOP Publishing},
volume = {2022},
number = {09},
pages = {039},
author = {Carron, Julien and Mirmelstein, Mark and Lewis, Antony},
title = {{CMB lensing from Planck PR4 maps}},
journal = {Journal of Cosmology and Astroparticle Physics}
}

@article{Qu_2024,
doi = {10.3847/1538-4357/acfe06},
url = {https://doi.org/10.3847/1538-4357/acfe06},
year = {2024},
month = {feb},
publisher = {The American Astronomical Society},
volume = {962},
number = {2},
pages = {112},
author = {Qu, Frank J. and Sherwin, Blake D. and Madhavacheril, Mathew S. and Han, Dongwon and Crowley, Kevin T. and Abril-Cabezas, Irene and Ade, Peter A. R. and Aiola, Simone and Alford, Tommy and Amiri, Mandana and Amodeo, Stefania and An, Rui and Atkins, Zachary and Austermann, Jason E. and Battaglia, Nicholas and Battistelli, Elia Stefano and Beall, James A. and Bean, Rachel and Beringue, Benjamin and Bhandarkar, Tanay and Biermann, Emily and Bolliet, Boris and Bond, J Richard and Cai, Hongbo and Calabrese, Erminia and Calafut, Victoria and Capalbo, Valentina and Carrero, Felipe and Carron, Julien and Challinor, Anthony and Chesmore, Grace E. and Cho, Hsiao-mei and Choi, Steve K. and Clark, Susan E. and Córdova Rosado, Rodrigo and Cothard, Nicholas F. and Coughlin, Kevin and Coulton, William and Dalal, Roohi and Darwish, Omar and Devlin, Mark J. and Dicker, Simon and Doze, Peter and Duell, Cody J. and Duff, Shannon M. and Duivenvoorden, Adriaan J. and Dunkley, Jo and Dünner, Rolando and Fanfani, Valentina and Fankhanel, Max and Farren, Gerrit and Ferraro, Simone and Freundt, Rodrigo and Fuzia, Brittany and Gallardo, Patricio A. and Garrido, Xavier and Gluscevic, Vera and Golec, Joseph E. and Guan, Yilun and Halpern, Mark and Harrison, Ian and Hasselfield, Matthew and Healy, Erin and Henderson, Shawn and Hensley, Brandon and Hervías-Caimapo, Carlos and Hill, J. Colin and Hilton, Gene C. and Hilton, Matt and Hincks, Adam D. and Hložek, Renée and Ho, Shuay-Pwu Patty and Huber, Zachary B. and Hubmayr, Johannes and Huffenberger, Kevin M. and Hughes, John P. and Irwin, Kent and Isopi, Giovanni and Jense, Hidde T. and Keller, Ben and Kim, Joshua and Knowles, Kenda and Koopman, Brian J. and Kosowsky, Arthur and Kramer, Darby and Kusiak, Aleksandra and La Posta, Adrien and Lague, Alex and Lakey, Victoria and Lee, Eunseong and Li, Zack and Li, Yaqiong and Limon, Michele and Lokken, Martine and Louis, Thibaut and Lungu, Marius and MacCrann, Niall and MacInnis, Amanda and Maldonado, Diego and Maldonado, Felipe and Mallaby-Kay, Maya and Marques, Gabriela A. and McMahon, Jeff and Mehta, Yogesh and Menanteau, Felipe and Moodley, Kavilan and Morris, Thomas W. and Mroczkowski, Tony and Naess, Sigurd and Namikawa, Toshiya and Nati, Federico and Newburgh, Laura and Nicola, Andrina and Niemack, Michael D. and Nolta, Michael R. and Orlowski-Scherer, John and Page, Lyman A. and Pandey, Shivam and Partridge, Bruce and Prince, Heather and Puddu, Roberto and Radiconi, Federico and Robertson, Naomi and Rojas, Felipe and Sakuma, Tai and Salatino, Maria and Schaan, Emmanuel and Schmitt, Benjamin L. and Sehgal, Neelima and Shaikh, Shabbir and Sierra, Carlos and Sievers, Jon and Sifón, Cristóbal and Simon, Sara and Sonka, Rita and Spergel, David N. and Staggs, Suzanne T. and Storer, Emilie and Switzer, Eric R. and Tampier, Niklas and Thornton, Robert and Trac, Hy and Treu, Jesse and Tucker, Carole and Ullom, Joel and Vale, Leila R. and Van Engelen, Alexander and Van Lanen, Jeff and van Marrewijk, Joshiwa and Vargas, Cristian and Vavagiakis, Eve M. and Wagoner, Kasey and Wang, Yuhan and Wenzl, Lukas and Wollack, Edward J. and Xu, Zhilei and Zago, Fernando and Zheng, Kaiwen},
title = {{The Atacama Cosmology Telescope: A Measurement of the DR6 CMB Lensing Power Spectrum and Its Implications for Structure Growth}},
journal = {The Astrophysical Journal}
}

@article{Madhavacheril_2024,
doi = {10.3847/1538-4357/acff5f},
url = {https://doi.org/10.3847/1538-4357/acff5f},
year = {2024},
month = {feb},
publisher = {The American Astronomical Society},
volume = {962},
number = {2},
pages = {113},
author = {Madhavacheril, Mathew S. and Qu, Frank J. and Sherwin, Blake D. and MacCrann, Niall and Li, Yaqiong and Abril-Cabezas, Irene and Ade, Peter A. R. and Aiola, Simone and Alford, Tommy and Amiri, Mandana and Amodeo, Stefania and An, Rui and Atkins, Zachary and Austermann, Jason E. and Battaglia, Nicholas and Battistelli, Elia Stefano and Beall, James A. and Bean, Rachel and Beringue, Benjamin and Bhandarkar, Tanay and Biermann, Emily and Bolliet, Boris and Bond, J Richard and Cai, Hongbo and Calabrese, Erminia and Calafut, Victoria and Capalbo, Valentina and Carrero, Felipe and Challinor, Anthony and Chesmore, Grace E. and Cho, Hsiao-mei and Choi, Steve K. and Clark, Susan E. and Córdova Rosado, Rodrigo and Cothard, Nicholas F. and Coughlin, Kevin and Coulton, William and Crowley, Kevin T. and Dalal, Roohi and Darwish, Omar and Devlin, Mark J. and Dicker, Simon and Doze, Peter and Duell, Cody J. and Duff, Shannon M. and Duivenvoorden, Adriaan J. and Dunkley, Jo and Dünner, Rolando and Fanfani, Valentina and Fankhanel, Max and Farren, Gerrit and Ferraro, Simone and Freundt, Rodrigo and Fuzia, Brittany and Gallardo, Patricio A. and Garrido, Xavier and Givans, Jahmour and Gluscevic, Vera and Golec, Joseph E. and Guan, Yilun and Hall, Kirsten R. and Halpern, Mark and Han, Dongwon and Harrison, Ian and Hasselfield, Matthew and Healy, Erin and Henderson, Shawn and Hensley, Brandon and Hervías-Caimapo, Carlos and Hill, J. Colin and Hilton, Gene C. and Hilton, Matt and Hincks, Adam D. and Hložek, Renée and Ho, Shuay-Pwu Patty and Huber, Zachary B. and Hubmayr, Johannes and Huffenberger, Kevin M. and Hughes, John P. and Irwin, Kent and Isopi, Giovanni and Jense, Hidde T. and Keller, Ben and Kim, Joshua and Knowles, Kenda and Koopman, Brian J. and Kosowsky, Arthur and Kramer, Darby and Kusiak, Aleksandra and La Posta, Adrien and Lague, Alex and Lakey, Victoria and Lee, Eunseong and Li, Zack and Limon, Michele and Lokken, Martine and Louis, Thibaut and Lungu, Marius and MacInnis, Amanda and Maldonado, Diego and Maldonado, Felipe and Mallaby-Kay, Maya and Marques, Gabriela A. and McMahon, Jeff and Mehta, Yogesh and Menanteau, Felipe and Moodley, Kavilan and Morris, Thomas W. and Mroczkowski, Tony and Naess, Sigurd and Namikawa, Toshiya and Nati, Federico and Newburgh, Laura and Nicola, Andrina and Niemack, Michael D. and Nolta, Michael R. and Orlowski-Scherer, John and Page, Lyman A. and Pandey, Shivam and Partridge, Bruce and Prince, Heather and Puddu, Roberto and Radiconi, Federico and Robertson, Naomi and Rojas, Felipe and Sakuma, Tai and Salatino, Maria and Schaan, Emmanuel and Schmitt, Benjamin L. and Sehgal, Neelima and Shaikh, Shabbir and Sierra, Carlos and Sievers, Jon and Sifón, Cristóbal and Simon, Sara and Sonka, Rita and Spergel, David N. and Staggs, Suzanne T. and Storer, Emilie and Switzer, Eric R. and Tampier, Niklas and Thornton, Robert and Trac, Hy and Treu, Jesse and Tucker, Carole and Ullom, Joel and Vale, Leila R. and Van Engelen, Alexander and Van Lanen, Jeff and van Marrewijk, Joshiwa and Vargas, Cristian and Vavagiakis, Eve M. and Wagoner, Kasey and Wang, Yuhan and Wenzl, Lukas and Wollack, Edward J. and Xu, Zhilei and Zago, Fernando and Zheng, Kaiwen},
title = {{The Atacama Cosmology Telescope: DR6 Gravitational Lensing Map and Cosmological Parameters}},
journal = {The Astrophysical Journal}
}

@article{DESI:2025zgx,
    author = "Karim, M. Abdul and others",
    collaboration = "DESI",
    title = "{DESI DR2 Results II: Measurements of Baryon Acoustic Oscillations and Cosmological Constraints}",
    eprint = "2503.14738",
    archivePrefix = "arXiv",
    primaryClass = "astro-ph.CO",
    month = "3",
    year = "2025"
}

@article{Brout:2022vxf,
    author = "Brout, Dillon and others",
    title = "{The Pantheon+ Analysis: Cosmological Constraints}",
    eprint = "2202.04077",
    archivePrefix = "arXiv",
    primaryClass = "astro-ph.CO",
    doi = "10.3847/1538-4357/ac8e04",
    journal = "Astrophys. J.",
    volume = "938",
    number = "2",
    pages = "110",
    year = "2022"
}

@article{Brout_2019,
doi = {10.3847/1538-4357/ab06c1},
url = {https://doi.org/10.3847/1538-4357/ab06c1},
year = {2019},
month = {mar},
publisher = {The American Astronomical Society},
volume = {874},
number = {1},
pages = {106},
author = {Brout, D. and Sako, M. and Scolnic, D. and Kessler, R. and D’Andrea, C. B. and Davis, T. M. and Hinton, S. R. and Kim, A. G. and Lasker, J. and Macaulay, E. and Möller, A. and Nichol, R. C. and Smith, M. and Sullivan, M. and Wolf, R. C. and Allam, S. and Bassett, B. A. and Brown, P. and Castander, F. J. and Childress, M. and Foley, R. J. and Galbany, L. and Herner, K. and Kasai, E. and March, M. and Morganson, E. and Nugent, P. and Pan, Y.-C. and Thomas, R. C. and Tucker, B. E. and Wester, W. and Abbott, T. M. C. and Annis, J. and Avila, S. and Bertin, E. and Brooks, D. and Burke, D. L. and Rosell, A. Carnero and Kind, M. Carrasco and Carretero, J. and Crocce, M. and Cunha, C. E. and Costa, L. N. da and Davis, C. and Vicente, J. De and Desai, S. and Diehl, H. T. and Doel, P. and Eifler, T. F. and Flaugher, B. and Fosalba, P. and Frieman, J. and García-Bellido, J. and Gaztanaga, E. and Gerdes, D. W. and Goldstein, D. A. and Gruen, D. and Gruendl, R. A. and Gschwend, J. and Gutierrez, G. and Hartley, W. G. and Hollowood, D. L. and Honscheid, K. and James, D. J. and Kuehn, K. and Kuropatkin, N. and Lahav, O. and Li, T. S. and Lima, M. and Marshall, J. L. and Martini, P. and Miquel, R. and Nord, B. and Plazas, A. A. and Roodman, A. and Rykoff, E. S. and Sanchez, E. and Scarpine, V. and Schindler, R. and Schubnell, M. and Serrano, S. and Sevilla-Noarbe, I. and Soares-Santos, M. and Sobreira, F. and Suchyta, E. and Swanson, M. E. C. and Tarle, G. and Thomas, D. and Tucker, D. L. and Walker, A. R. and Yanny, B. and Zhang, Y. and (DES COLLABORATION)},
title = {{First Cosmology Results Using Type Ia Supernovae from the Dark Energy Survey: Photometric Pipeline and Light-curve Data Release}},
journal = {The Astrophysical Journal}
}

@article{DES:2024tys,
    author = "Abbott, T. M. C. and others",
    collaboration = "DES",
    title = "{The Dark Energy Survey: Cosmology Results With $\sim$1500 New High-redshift Type Ia Supernovae Using The Full 5-year Dataset}",
    eprint = "2401.02929",
    archivePrefix = "arXiv",
    primaryClass = "astro-ph.CO",
    reportNumber = "FERMILAB-PUB-23-0821-PPD, DES-2023-805",
    month = "1",
    year = "2024"
}

@article{Lewis:1999bs,
      author         = "Lewis, Antony and Challinor, Anthony and Lasenby,
                        Anthony",
      title          = "{Efficient computation of CMB anisotropies in closed FRW
                        models}",
      journal        = "Astrophys. J.",
      volume         = "538",
      year           = "2000",
      pages          = "473-476",
      doi            = "10.1086/309179",
      eprint         = "astro-ph/9911177",
      archivePrefix  = "arXiv",
      primaryClass   = "astro-ph",
      SLACcitation   = "%%CITATION = ASTRO-PH/9911177;%%",
      url            = {https://arxiv.org/abs/astro-ph/9911177}
}

@article{Howlett:2012mh,
      author         = "Howlett, Cullan and Lewis, Antony and Hall, Alex and
                        Challinor, Anthony",
      title          = "{CMB power spectrum parameter degeneracies in the era of
                        precision cosmology}",
      journal        = "JCAP",
      volume         = "1204",
      year           = "2012",
      pages          = "027",
      doi            = "10.1088/1475-7516/2012/04/027",
      eprint         = "1201.3654",
      archivePrefix  = "arXiv",
      primaryClass   = "astro-ph.CO",
      SLACcitation   = "%%CITATION = ARXIV:1201.3654;%%",
      url            = {https://arxiv.org/abs/1201.3654}
}

@article{Neal:2005,
      author    = {{Neal}, R.~M.},
      title     = "{Taking Bigger Metropolis Steps by Dragging Fast Variables}",
      journal   = {ArXiv Mathematics e-prints},
      eprint    = {math/0502099},
      year      = 2005,
      month     = feb,
      adsurl    = {http://adsabs.harvard.edu/abs/2005math......2099N},
      url       = {https://arxiv.org/abs/math/0502099}
}

@article{Torrado:2020dgo,
    author = "Torrado, Jesus and Lewis, Antony",
    title = "{Cobaya: Code for Bayesian Analysis of hierarchical physical models}",
    eprint = "2005.05290",
    archivePrefix = "arXiv",
    primaryClass = "astro-ph.IM",
    reportNumber = "TTK-20-15",
    doi = "10.1088/1475-7516/2021/05/057",
    journal = "JCAP",
    volume = "05",
    pages = "057",
    year = "2021"
}

@article{Lewis_2025,
doi = {10.1088/1475-7516/2025/08/025},
url = {https://doi.org/10.1088/1475-7516/2025/08/025},
year = {2025},
month = {aug},
publisher = {IOP Publishing},
volume = {2025},
number = {08},
pages = {025},
author = {Lewis, Antony},
title = {{GetDist: a Python package for analysing Monte Carlo samples}},
journal = {Journal of Cosmology and Astroparticle Physics}
}

@article{GelmanRubin,
author = {Andrew Gelman and Donald B. Rubin},
title = {{Inference from Iterative Simulation Using Multiple Sequences}},
volume = {7},
journal = {Statistical Science},
number = {4},
publisher = {Institute of Mathematical Statistics},
pages = {457 -- 472},
year = {1992},
doi = {10.1214/ss/1177011136},
URL = {https://doi.org/10.1214/ss/1177011136}
}

@article{wang,
  title = {Figure of merit for dark energy constraints from current observational data},
  author = {Wang, Yun},
  journal = {Phys. Rev. D},
  volume = {77},
  issue = {12},
  pages = {123525},
  numpages = {7},
  year = {2008},
  month = {Jun},
  publisher = {American Physical Society},
  doi = {10.1103/PhysRevD.77.123525},
  url = {https://link.aps.org/doi/10.1103/PhysRevD.77.123525}
}

@article{SixiangWen_2018,
doi = {10.1088/1475-7516/2018/07/011},
url = {https://doi.org/10.1088/1475-7516/2018/07/011},
year = {2018},
month = {jul},
publisher = {},
volume = {2018},
number = {07},
pages = {011},
author = {Wen, Sixiang and Wang, Shuang and Luo, Xiaolin},
title = {Comparing dark energy models with current observational data},
journal = {Journal of Cosmology and Astroparticle Physics}
}

@Article{Hu2014,
author={Hu, YaZhou
and Li, Miao
and Li, XiaoDong
and Zhang, ZhenHui},
title={Investigating the possibility of a turning point in the dark energy equation of state},
journal={Science China Physics, Mechanics {\&} Astronomy},
year={2014},
month={Aug},
day={01},
volume={57},
number={8},
pages={1607-1612},
issn={1869-1927},
doi={10.1007/s11433-014-5497-y},
url={https://doi.org/10.1007/s11433-014-5497-y}
}
